\documentclass[twocolumn]{emulateapj}
\usepackage{color}

\usepackage[usenames,dvipsnames,svgnames,table]{xcolor}
\usepackage[colorlinks=true,
           linkcolor=blue,
           urlcolor=blue,
           citecolor=blue]{hyperref}

\begin{document}

\title{X-ray and EUV Observations of Simultaneous Short and Long Period Oscillations in Hot Coronal Arcade Loops}

\author{PANKAJ KUMAR\altaffilmark{1}, VALERY M. NAKARIAKOV\altaffilmark{2,3,4}, KYUNG-SUK CHO\altaffilmark{1}}
\affil{$^1$Korea Astronomy and Space Science Institute (KASI), Daejeon, 305-348, Republic of Korea}
\affil{$^2$Centre for Fusion, Space and Astrophysics, Department of Physics, University of Warwick, CV4 7AL, UK}
\affil{$^3$School of Space Research, Kyung Hee University, Yongin, 446-701, Gyeonggi, Republic of Korea}
\affil{$^4$ Central Astronomical Observatory at Pulkovo of RAS, 196140 St Petersburg, Russia}
\email{pankaj@kasi.re.kr}

\begin{abstract}
We report decaying quasi-periodic intensity oscillations in the X-ray (6--12~keV) and extreme ultraviolet (EUV) channels (131, 94, 1600, 304~\AA) observed by the Fermi GBM (Gamma-ray Burst Monitor) and SDO/AIA ({\it Solar Dynamic Observatory/Atmospheric Imaging Assembly}), respectively, during a C-class flare. The estimated period of oscillation and decay time in the X-ray channel (6--12~keV) was about 202~s and 154~s, respectively. A similar oscillation period was detected at the footpoint of the arcade loops in the AIA 1600 and 304~\AA~ channels. Simultaneously, AIA hot channels (94 and 131~\AA) reveal propagating EUV disturbances bouncing back and forth between the footpoints of the arcade loops. The period of the oscillation and decay time were about 409~s and 1,121~s, respectively. The characteristic phase speed of the wave is about 560~km~s$^{-1}$ for  about 115~Mm loop length, which is roughly consistent with the sound speed at the temperature about 10--16~MK (480--608~km~s$^{-1}$). 
These EUV oscillations are consistent with the SOHO/SUMER ({\it Solar and Heliospheric Observatory/Solar Ultraviolet Measurement of Emitted Radiation}) Doppler-shift oscillations interpreted as the global standing slow magnetoacoustic wave excited by a flare. 
The flare occurred at one of the footpoints of the arcade loops, where
the magnetic topology was a 3D fan-spine with a null-point. Repetitive reconnection at this footpoint could cause the periodic acceleration of non-thermal electrons that propagated to the opposite footpoint along the arcade
and precipitating there, causing the observed 202-s periodicity.
Other possible interpretations, e.g. the second harmonics of the slow mode are also discussed.   
 \end{abstract}

\section{INTRODUCTION}
The study of magnetohydrodynamic (MHD) waves and oscillations in the solar atmosphere is very important for understanding the energy release processes, particle acceleration or heating mechanisms and to determine the plasma parameters indirectly by coronal seismology \citep{roberts2000,demoortel2012,liu2014}.

Standing slow-mode oscillations in coronal loops have been discovered by the Solar Ultraviolet Measurements of Emitted Radiation (SUMER) spectrograph onboard the Solar and Heliospheric Observatory (SOHO), by recording the Doppler-shift and intensity measurements of hot coronal lines (i.e., \ion{Fe}{19} and \ion{Fe}{21} lines, with the formation temperature greater than 6~MK) \citep{kliem2002,wang2002,wang2003a,wang2003b,wang2011}. The statistical study of a large set of events reveal the period of oscillation of 7--31~minutes and initial Doppler speed of about 200~km~s$^{-1}$ \citep{wang2003b}. Similar Doppler-shift oscillations have been detected by YOHKOH/BCS \citep{mariska2005,mariska2006,mariska2008}. These longitudinal oscillations are of compressive nature.  The observed strong damping is usually interpreted in terms of thermal conduction \citep{ofman2002}. Several authors included other physical effects (e.g., viscosity, geometry, stratification, nonlinear steepening, mode coupling, shock dissipation, wave leakage etc.) in order to explain the observational damping time of the oscillations \citep{nakariakov2004,tsiklauri2004,taroyan2005,selwa2005,selwa2007,pandey2006,bradshaw2008,erdelyi2008,haynes2008,verwichte2008,ogro2009,selwa2009, grus2011}. 

Excitation of slow magnetoacoustic oscillations in hot coronal loops has been intensively studied theoretically too.  One-dimensional (1D) MHD numerical simulation performed by \citet{nakariakov2004} and \citet{tsiklauri2004} showed the efficient generation of fundamental and second harmonic modes by a localized energy release at one or both footpoints of the loop. \citet{selwa2005} showed the excitation of a fundamental mode by launching a pressure pulse at one of the footpoints of the loop. Using a 3D loop model, \citet{selwa2009} demonstrated that a fast pulse at one of the footpoints of the loop causes a slow mode wave bouncing between the footpoints. However, the  excitation of these oscillation and decaying mechanisms are still insufficiently understood and remain a subject of further investigations. 

At present, high temporal and spatial resolution observations from SDO/AIA are very useful to study the waves and oscillations in the solar corona, in particular the slow modes. \citet{kim2012} reported the first simultaneous observations of the plasma density and EUV intensity oscillations with the period of about 12.6~minutes, using Nobeyama 17~GHz and AIA 335~\AA~ channels, respectively. The observed properties of these oscillations match the SUMER Dopper-shift oscillations associated with the slow magnetoacoustic mode. Recently, \citet{kumar2013w} discovered the first direct observation of a compressive longitudinal wave reflecting back and forth between the footpoints of a loop system during a C-class flare, which occurred at one of the footpoints of the loop system. The longitudinal wave was observed only in the hot-temperature AIA channels, i.e., 131 and 94~\AA~ (at the temperature in the range of 8--10~MK), thus consistent with the SUMER hot loop oscillations and with their interpretation as a slow-mode wave.

Quasi-periodic pulsations (QPP) are often observed during solar flares (with the periods ranging from a fraction of a second to several minutes) at different wavelengths, i.e.,  X-ray, EUV, gamma-rays and radio \citep{asc2004,nakariakov2009,nakariakov2010}. Similar oscillations are also detected in stellar flares \citep{mathioudakis2003,pandey2009,anfinogentov2013}. Several mechanisms have been proposed to explain the QPPs in the hard X-ray and radio wavelengths. For example, a model of bursty magnetic reconnection demonstrates the formation and coalescence of multiple plasmoids during the tearing-mode instability at the flaring current-sheet \citep{kliem2000}. Tearing of current-sheet and coalescence of multiple plasmoids during magnetic reconnection can modulate particle acceleration observed in the radio and hard X-ray channels. Bidirectional plasmoids are generally observed during the bursty magnetic reconnection that generates a QPP of few seconds in the form of radio drifting pulsating structures \citep{barta2008, kumar2013p}. 
 The loop-loop interaction model \citep{tajima1987} have also shown the generation of QPP (in radio and hard X-ray) by coalescence or interaction of current-carrying loops \citep{kumar2010l}.

In addition, MHD waves can also trigger the periodic reconnection, therefore, causing the repeative acceleration of non-thermal particles, resulting in the QPP observed in the radio, EUV and X-ray wavelengths. \citet{ning2004} reported the bursts of explosive events with a period of 3--5~minutes, similar to the period of chromosphere and transition-region oscillations. They suggested that the periodic reconnection could be triggered by a compressive or torsional Alfv\'en wave. 
\citet{doyle2006} reported repetitive explosive events at a coronal-hole boundary with a period of 3--5~minutes and interpreted them in terms of kink-mode of the flux-tube. \citet{chen2006} modeled the explosive events and showed that the slow-mode wave can trigger the periodic reconnection. A  model based on an external fast-magnetoacoustic oscillation, proposed by \citet{nakariakov2006}, shows that a transverse oscillation of a coronal loop situated near the magnetic reconnection site could periodically trigger flaring energy releases and hence produce QPP. Repeated reflection of the slow-mode wave from the footpoints of the arcade loops can generate QPP observed in a two ribbon flare \citep{nakariakov2011}. Although there are so many theoretical mechanisms to explain the QPPs, the exact mechanism should be investigated by analyzing high-resolution multi-wavelength observations.

In this paper, we analyze a rare event that shows almost simultaneous short period (202~s) and long period (409~s) decaying oscillations in the X-ray and EUV channels, observed by the Fermi GBM and SDO/AIA during a C-class flare on 20 July 2013. The paper includes high resolution observations from the SDO/AIA and Hinode, which are different 
from SUMER hot loop observations obtained off-limb at a fixed slit position without the information about the magnetic field configuration. The magnetic configuration of the flare site revealed flux emergence, showing the fan-spine topology with a null-point at the boundary of the active region NOAA 11793. In Section~2, we present the observations and results. In the last section, we summarize and discuss the results.

\section{OBSERVATIONS AND RESULTS}

The {\it Atmospheric Image Assembly} (AIA; \citealt{lemen2012}) onboard the {\it Solar Dynamics Observatory}  
(SDO; \citealt{pesnell2012}) acquires full disk images of the Sun (field-of-view $\sim$1.3 $R_\odot$) with the spatial resolution of 1.5$\arcsec$ 
(0.6$\arcsec$~pixel$^{-1}$) 
and cadence of 12~sec, in ten extreme ultraviolet (EUV) and UV channels. This study utilizes 171~\AA\ (\ion{Fe}{9}, $T\approx$0.7 MK), 94~\AA\ (\ion{Fe}{18}, $T\approx$6.3~MK), 131~\AA~ (\ion{Fe}{8}, \ion{Fe}{21}, \ion{Fe}{23}, i.e., 0.4, 10, 16~MK, respectively), 304~\AA~(\ion{He}{2}, T$\approx$0.05~MK) and 1600~\AA\ (\ion{C}{4} + continuum, $T \approx 0.01$~MK) images. We also used Heliospheric and Magnetic
Imager (HMI) magnetograms \citep{schou2012} to investigate the magnetic configuration of the active region. 

The active region (AR) NOAA 11793 was located near the disk center (N21W02) on 20 July 2013, showing a  $\beta\gamma\delta$ magnetic configuration. The decaying oscillation reported here, was observed during a C2.1 class flare. The flare was triggered at the edge of the AR, where an anemone or fan shaped loops emerged within the preexisting active region. The flare started at $\sim$03:34~UT, peaked at $\sim$03:38~UT, and ended at $\sim$03:44~UT.

\begin{figure*}
\centering{
\includegraphics[width=12cm]{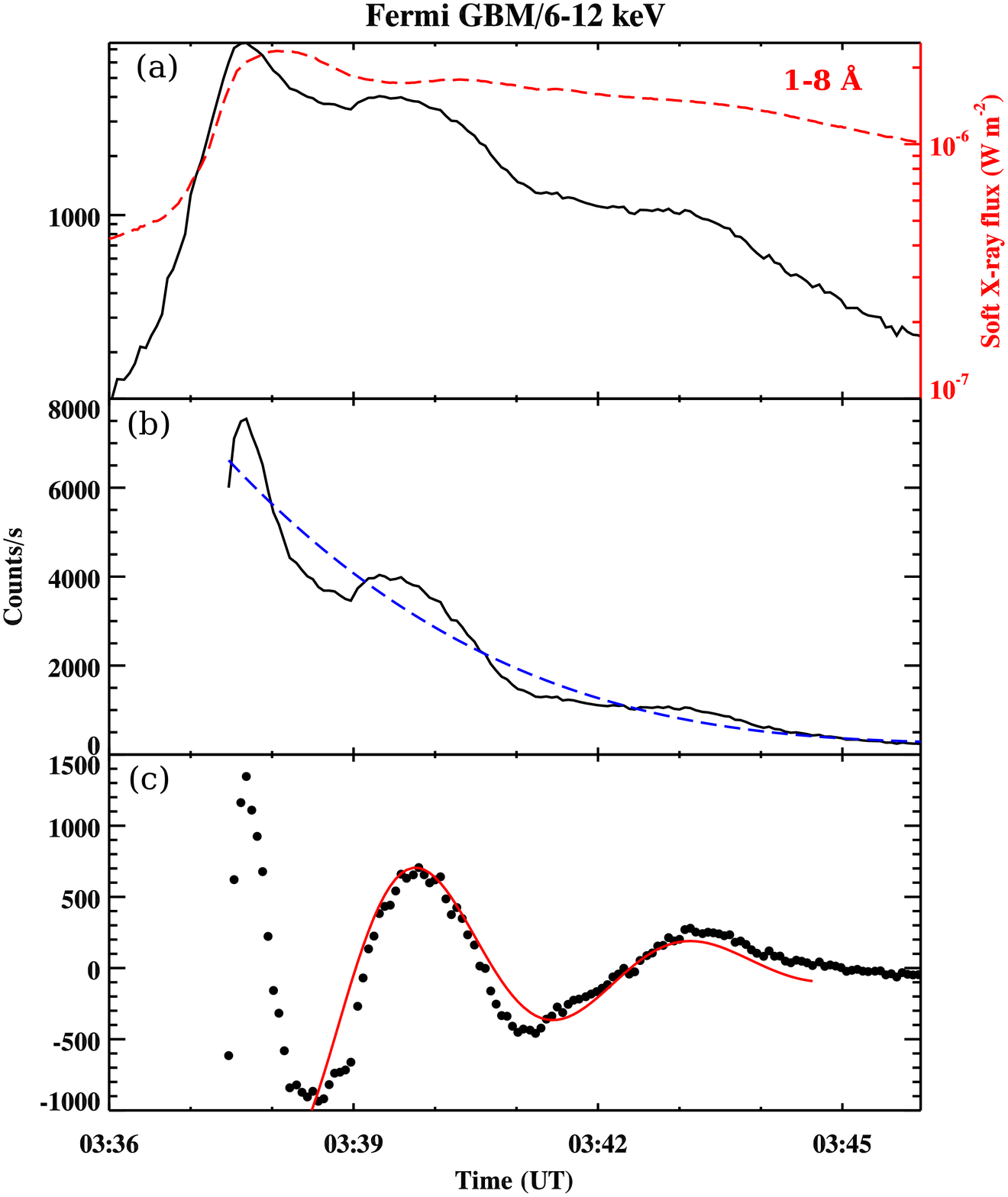}
}
\caption{(a) Fermi GBM flux profile of the flare in the 6--12 keV channel, and GOES soft X-ray flux profile (red, log scale) in the 1--8~\AA~ channel. (b) A cubic polynomial function (blue curve) is used to detrend the light curve (linear scale). (c) Red curve shows the best fitted decaying sine function.}
\label{fermi}
\end{figure*}


\subsection{Decaying Oscillation}

Figure \ref{fermi}a displays the Fermi GBM (Gamma-ray Burst Monitor) light curve in 6--12~keV channel, and the GOES soft X-ray profile (red) in the 1--8~\AA\ channel. The Fermi GBM light curve shows a clear decaying oscillation for the duration of $\sim$10~min. The oscillation was not observed in the higher energy channels ($>$12~keV). To extract the oscillation profile, we detrended the light curve by subtracting a cubic polynomial profile (blue curve in panel (b)) from the original light curve. The detrended light curve is shown in the bottom panel (c). We fitted a decaying sine function (using MPFIT) to determine the period and decay time of the oscillation, 

\begin{equation}
I(t)=A\mbox{ sin}(\frac{2\pi t }{P}+\phi)\mbox{ exp}(\frac{-t}{\tau}),
\label{eq1}
\end{equation}

\noindent
where $A$, $P$, $\tau$, and $\phi$ are the amplitude, period, decay time and initial phase, respectively. The best fitted curve is shown by the thick red curve in panel (c). The estimated period of the oscillation is $P \approx 202$~s, and the decay time  is $\tau \approx 154$~s.

\begin{figure*}
\centering{
\includegraphics[width=8cm]{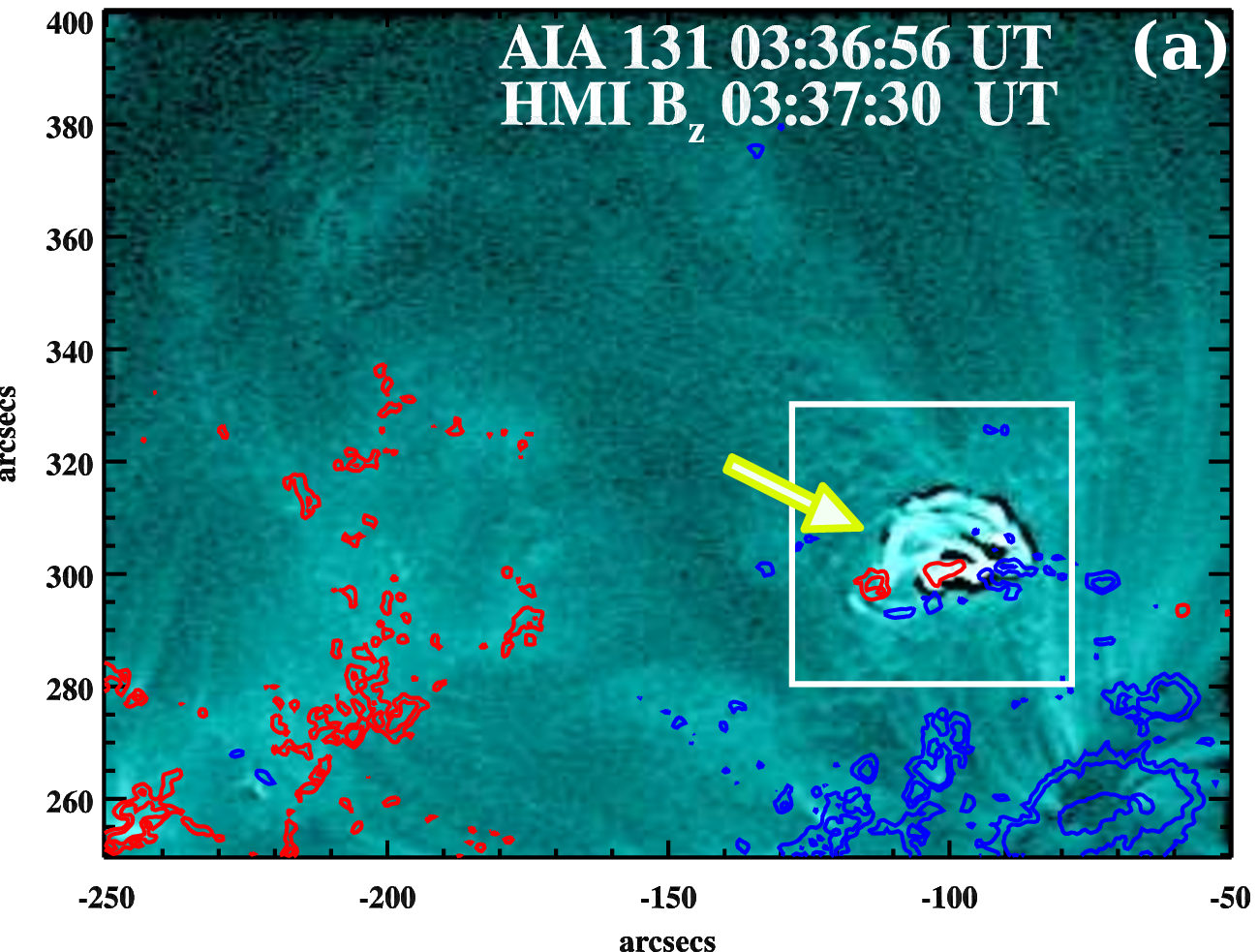}
\includegraphics[width=8cm]{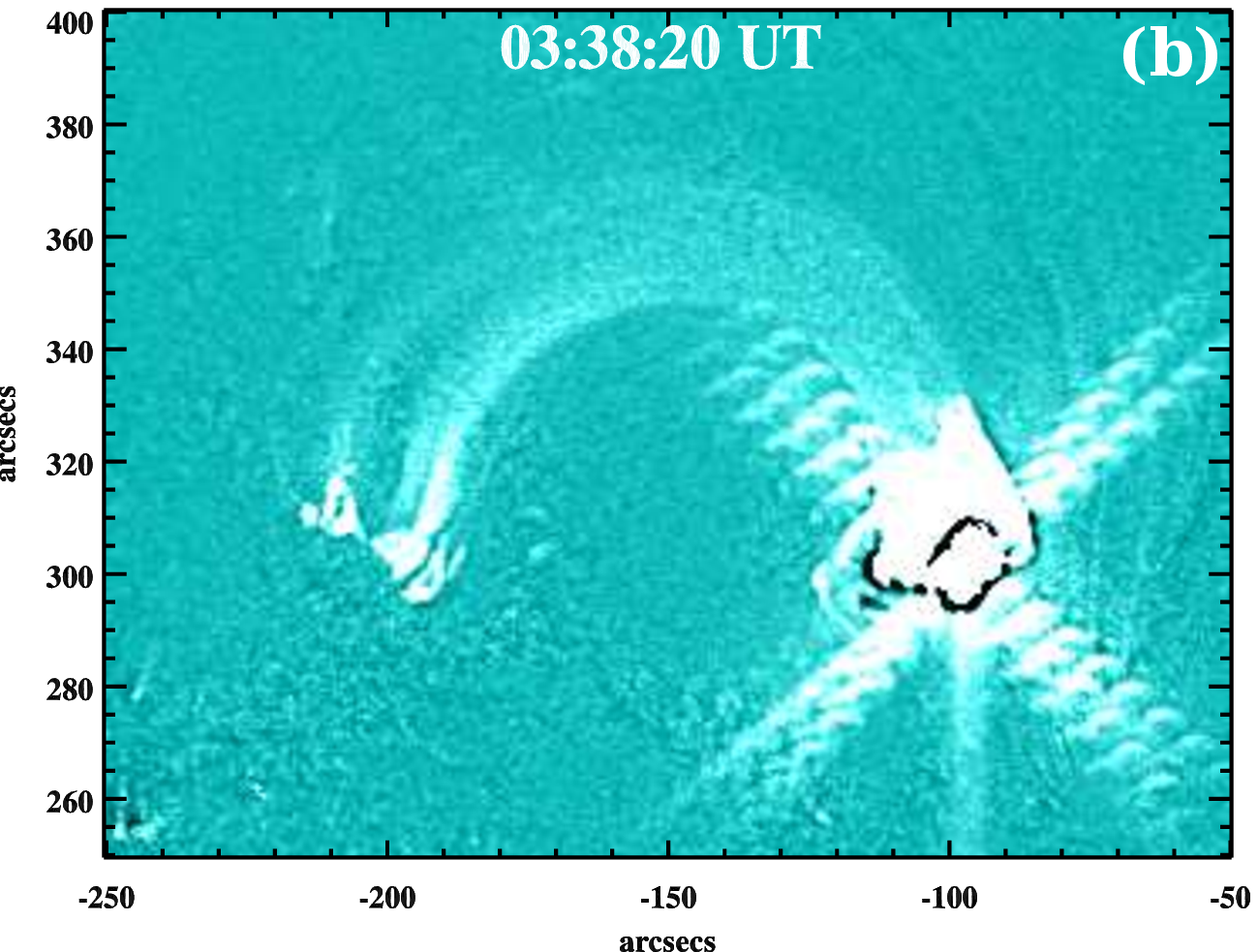}
\includegraphics[width=8cm]{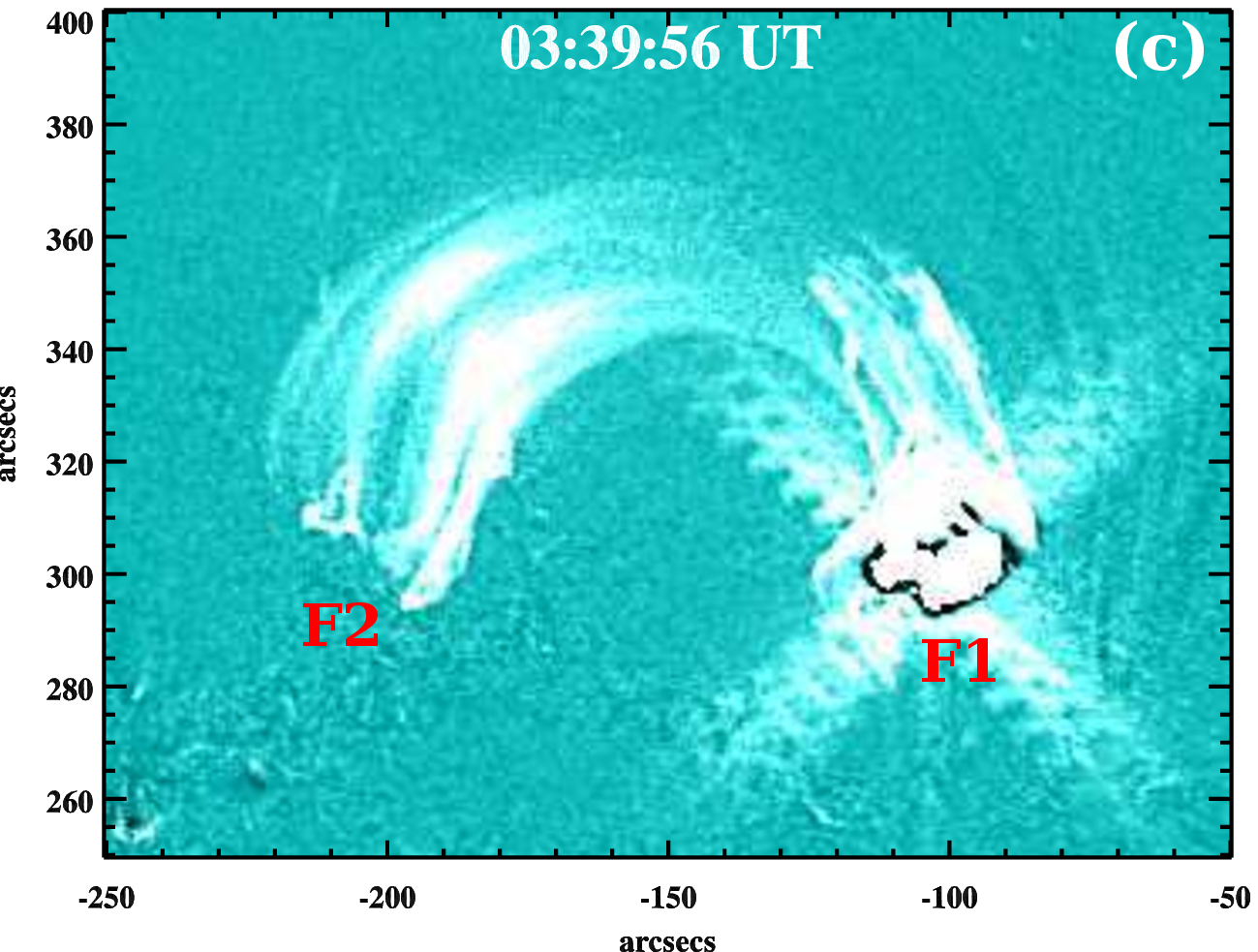}
\includegraphics[width=8cm]{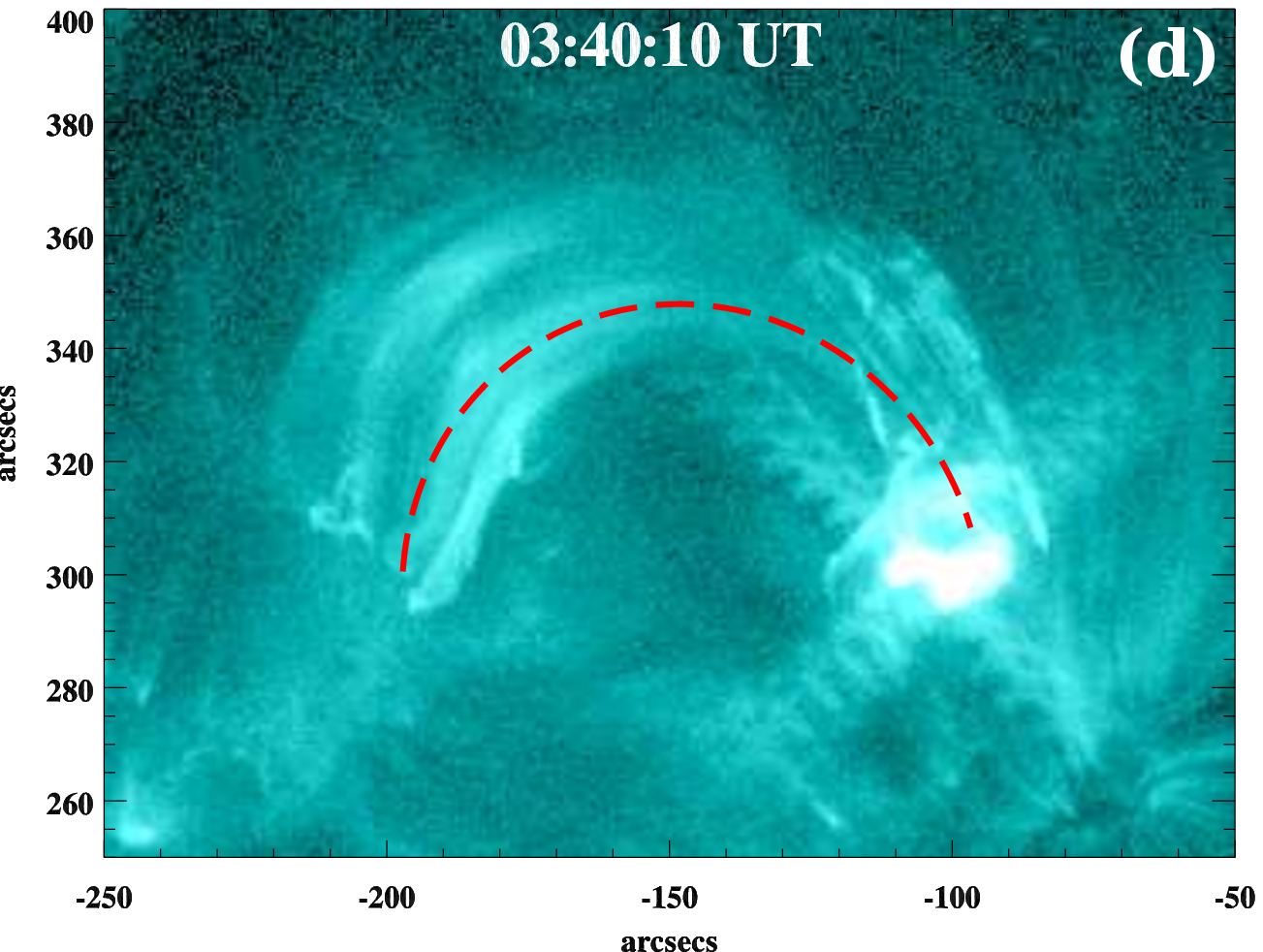}
\includegraphics[width=8cm]{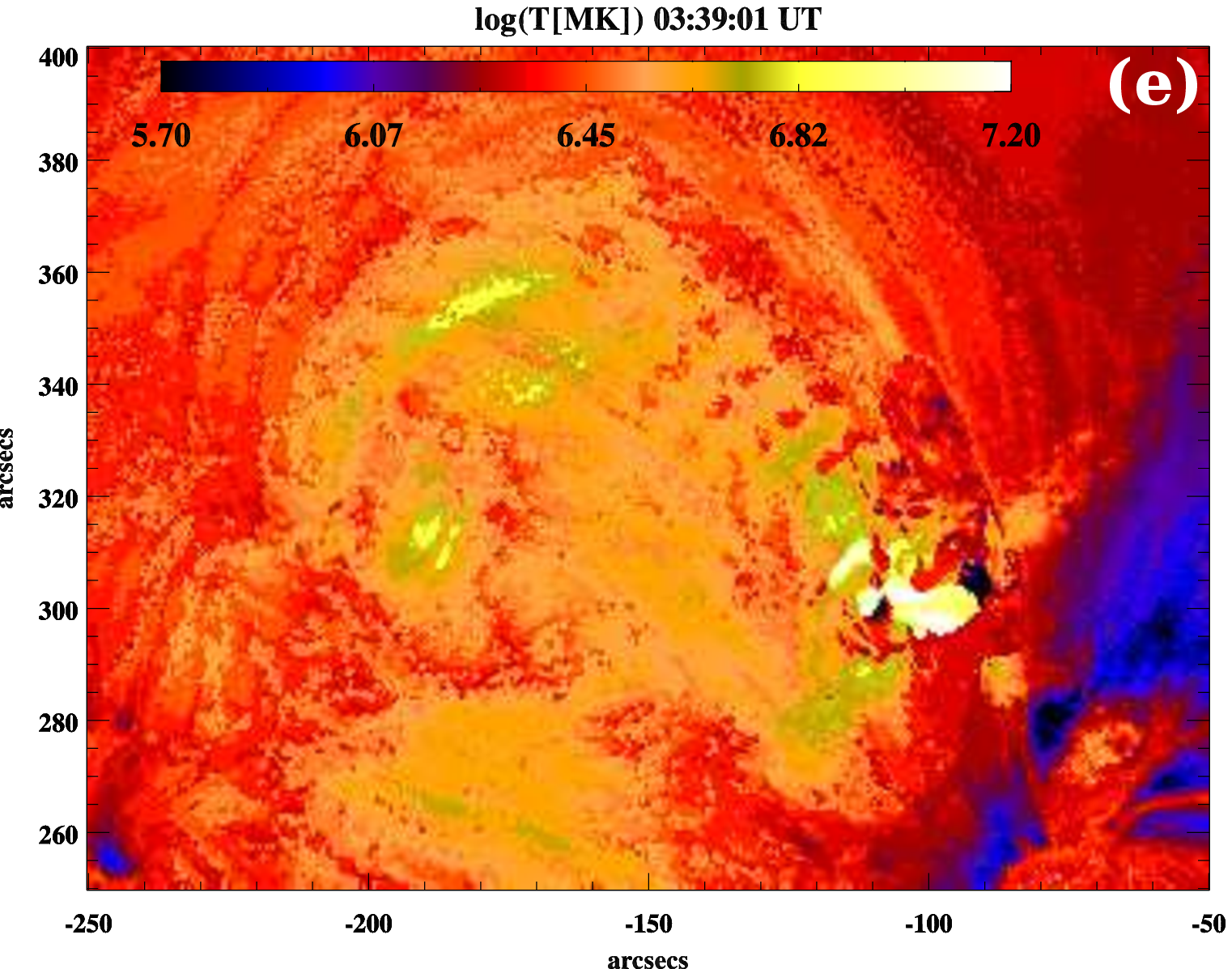}
\includegraphics[width=8cm]{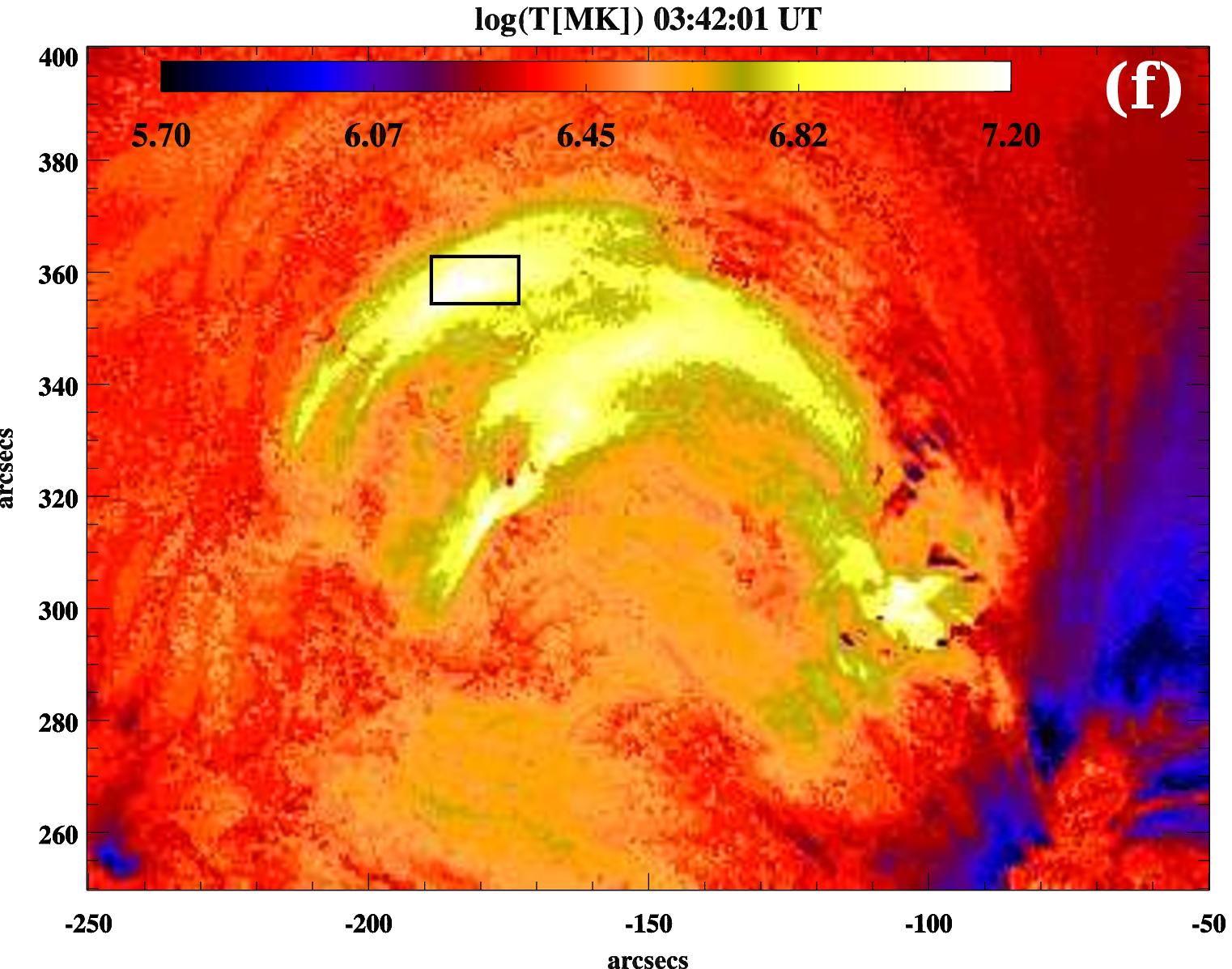}
}
\caption{(a-d) SDO/AIA 131~\AA~ images during the flare that occurred at one of the footpoints of the arcade loops. HMI magnetogram contours of positive (red) and negative (blue) polarities are overlaid on AIA 131~\AA~ image in the first panel. The contour levels are $\pm$400, $\pm$800, and $\pm$1600 Gauss. F1 and F2 represent the footpoints of the hot arcade loops. The flare was triggered at footpoint F1. The dashed red curve shows the semi-circular fit to the loop system. (e-f) DEM peak temperature maps derived from near simultaneously AIA six channel intensity images. The mean and maximum value of DEM peak temperature within the box region are 6.4~MK and 15.8~MK, respectively. }
\label{aia131}
\end{figure*}

\begin{figure*}
\centering{
\includegraphics[width=8cm]{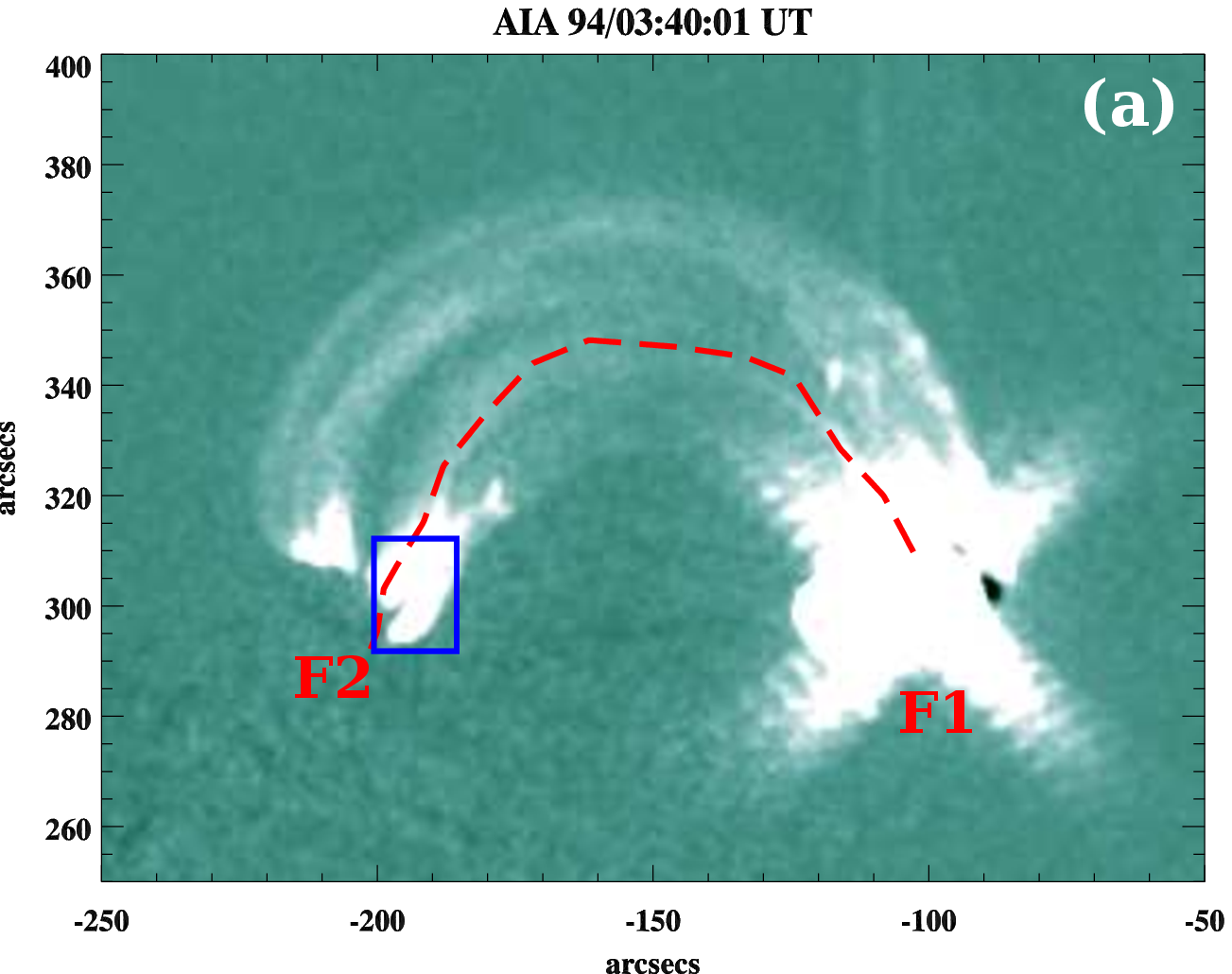}
\includegraphics[width=4cm]{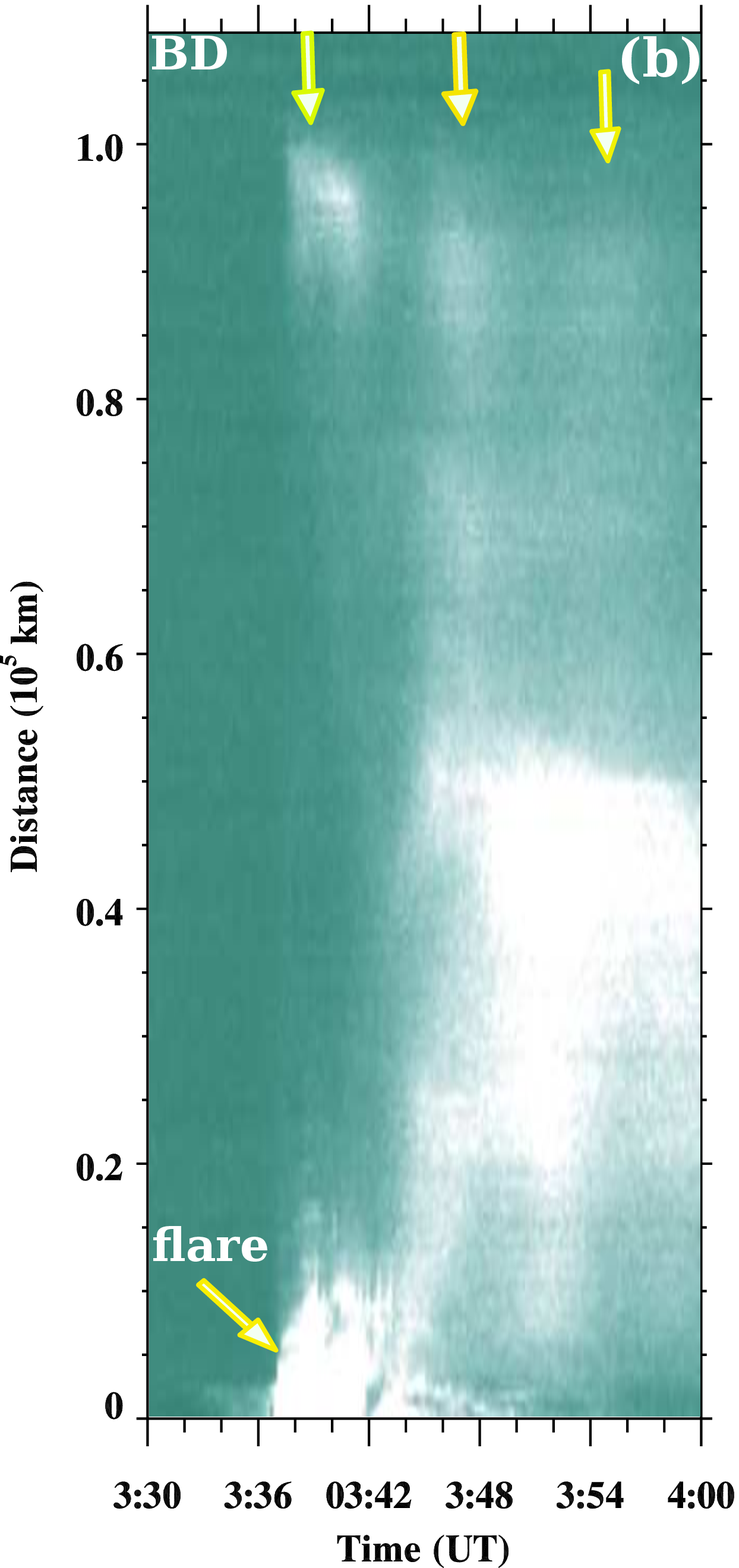}
\includegraphics[width=4cm]{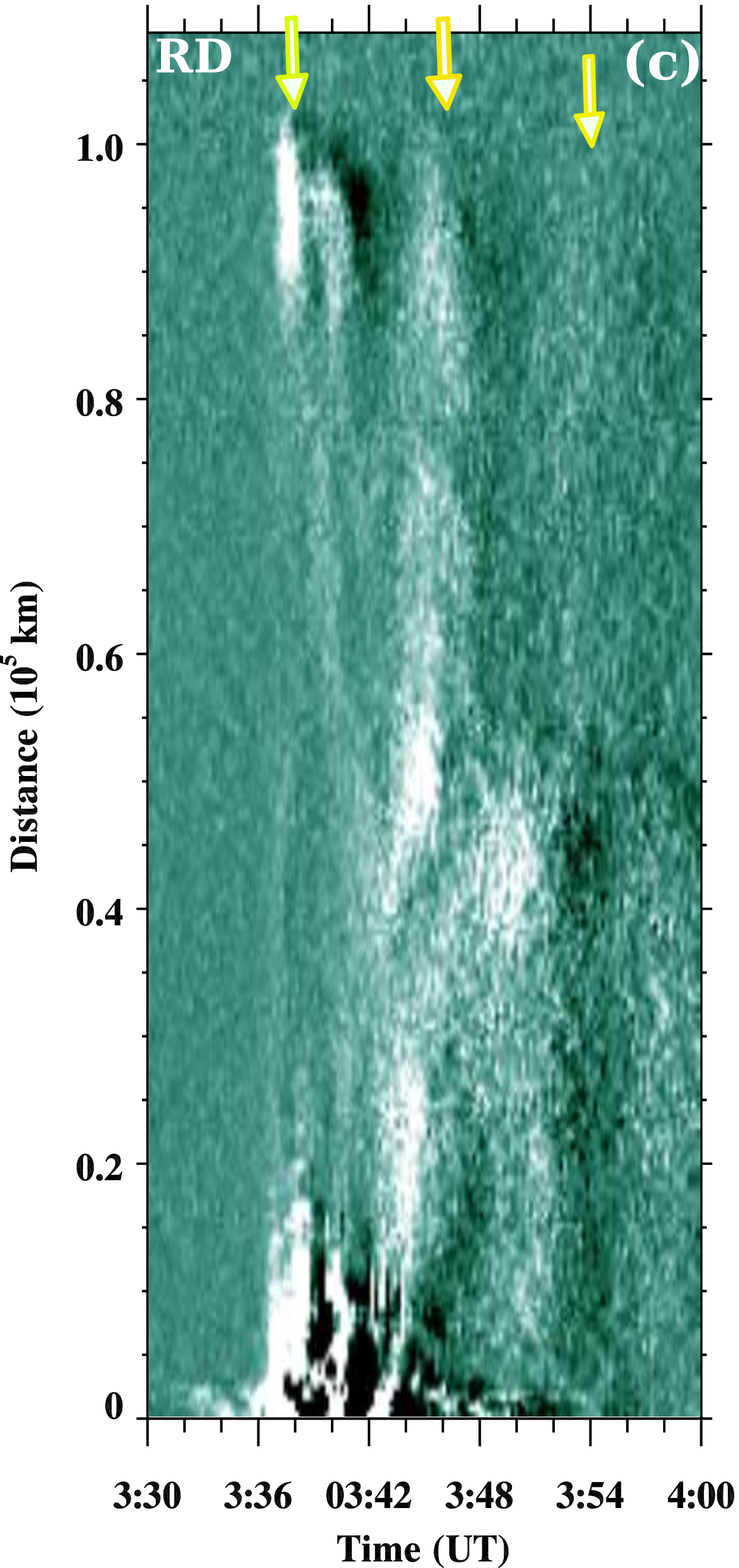}

\includegraphics[width=10cm]{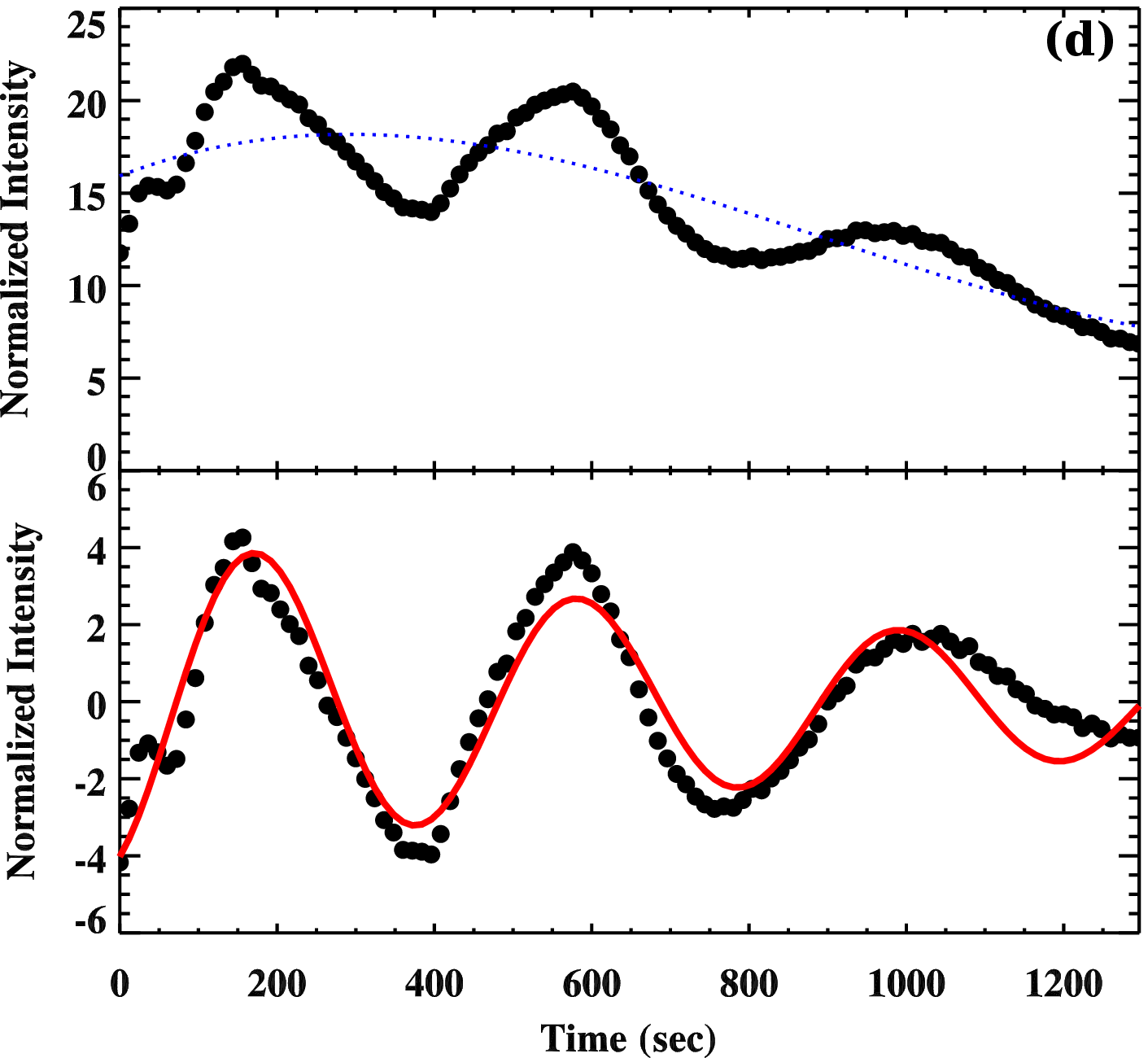}

}
\caption{(a-c) Distance-time plots of the intensity distribution (base difference and running difference, i.e., BD and RD) along the hot loop (red curve, panel a). (d) AIA 94~\AA~ base difference (mean) intensity plot extracted from the rectangular box (marked by blue color). The start time is $\sim$03:38~UT. Blue dashed line is a cubic polynomial trend. The best fitted (de-trended) light curve (thick red curve). The estimated period of the oscillation and decay time are 409 and 1121~s, respectively.}
\label{aia94}
\end{figure*}

Figure \ref{aia131} shows the AIA~131~\AA~ intensity (a, d) and base-difference (b, c) images of the flare site. Figure \ref{aia131}a is overlaid by HMI magnetogram contours of positive (red) and negative (blue) polarities. Emerging fan-shaped low-lying loops where the C2.1-class flare was triggered are marked by an arrow. These panels indicate that the flare occurred at one of the footpoints of the arcade loops. We could observe a reflecting EUV disturbance (observed only in the AIA hot channels, i.e., 131 and 94~\AA) that moves back and forth along the arcade loops (see AIA 94~\AA~ movie, i.e., aia94$\_$bd.mp4).

To determine the characteristics of the oscillation, we need to estimate the loop length and sound speed within the arcade loops.
For the loop length estimation, we fitted the data with a 3D semicircular loop model \citep{asc2002} shown by the red dotted curve (Figure \ref{aia131}d). In this model, we need to fit two free parameters: the loop center offset from the solar surface (h$_0$) and loop inclination angle ($\theta$) from the vertical. The best fitted parameters for the loop are, $h_0$=-0.01~$R_\odot$ (solar radii), and $\theta =-50^\circ$. The estimated loop length is about 158$\arcsec$.

We used an automatic code developed by \citet{asc2013} to determine the differential emission measure (DEM) distribution (peak temperature) of the flare site and arcade loops. This code provides the DEM peak temperature $T_{p}$, emission measure $EM_{p}$, and sigma $\sigma _{p}$. Figure \ref{aia131}e-f show the peak temperature maps ($\log T_p$, MK) of the flare region and arcade loops. The  arcade loops were visible only in the AIA hot channels (131 and 94~\AA), therefore, the plasma temperature of the arcade loops is most likely between 6.3--16~MK. We choose a temperature range of 0.5--16~MK in the automatic DEM code. The peak temperature of the flare site was about 10~MK. For the arcade loops, the average and maximum values of the peak temperature within the box region are 6.4 and 15.8~MK, respectively. Using these temperatures, we can estimate the values of the sound speed, as $c_{s}=152\sqrt{T}$ where the speed is in km~s$^{-1}$ and temperature is in MK. Thus, we obtain 385~km~s$^{-1}$ and 605~km~s$^{-1}$, respectively.

Apart from the decaying oscillation in the 6--12~keV, we also observed a periodic EUV disturbance.
AIA 131 and 94~\AA\ images reveal the EUV disturbances moving back and forth along the arcade loops. These disturbances were observed only in the AIA hot channels. To investigate the details of these periodic disturbances, we extracted the intensity by selecting a path along the arcade loop (marked by the red dotted curve) using AIA 94~\AA~ images (Figure \ref{aia94}). The distance-time plot for the base and running difference intensity along the selected path are shown in Figure \ref{aia94}(b-c). Both panels clearly show an EUV disturbance reflecting back and forth between the footpoints of the arcade loops (marked by arrows). To determine the period of oscillation, we selected a rectangular box (blue) at the footpoint of the arcade loop. We extracted the average intensity within the box using AIA 94~\AA~ base difference images. The intensity profile reveals a decaying oscillation in the original light curve. We detrended the light curve by subtracting a cubic polynomial function (blue dotted curve). We fitted an exponential decaying sine function (\ref{eq1}). The best fitted curve is shown by a thick red curve. The period of the oscillation and the damping time are $\sim$409~s and 1121~s, respectively.
Using this period, the phase speed of the wave is $2L/P\approx 560$~km~s$^{-1}$.
 Considering the uncertainties of about 10-15$\%$ in the DEM peak temperature and loop-length estimation, the phase speed roughly matches the sound speed in the hot arcade loops.

\begin{figure*}
\centering{
\includegraphics[width=8.0cm]{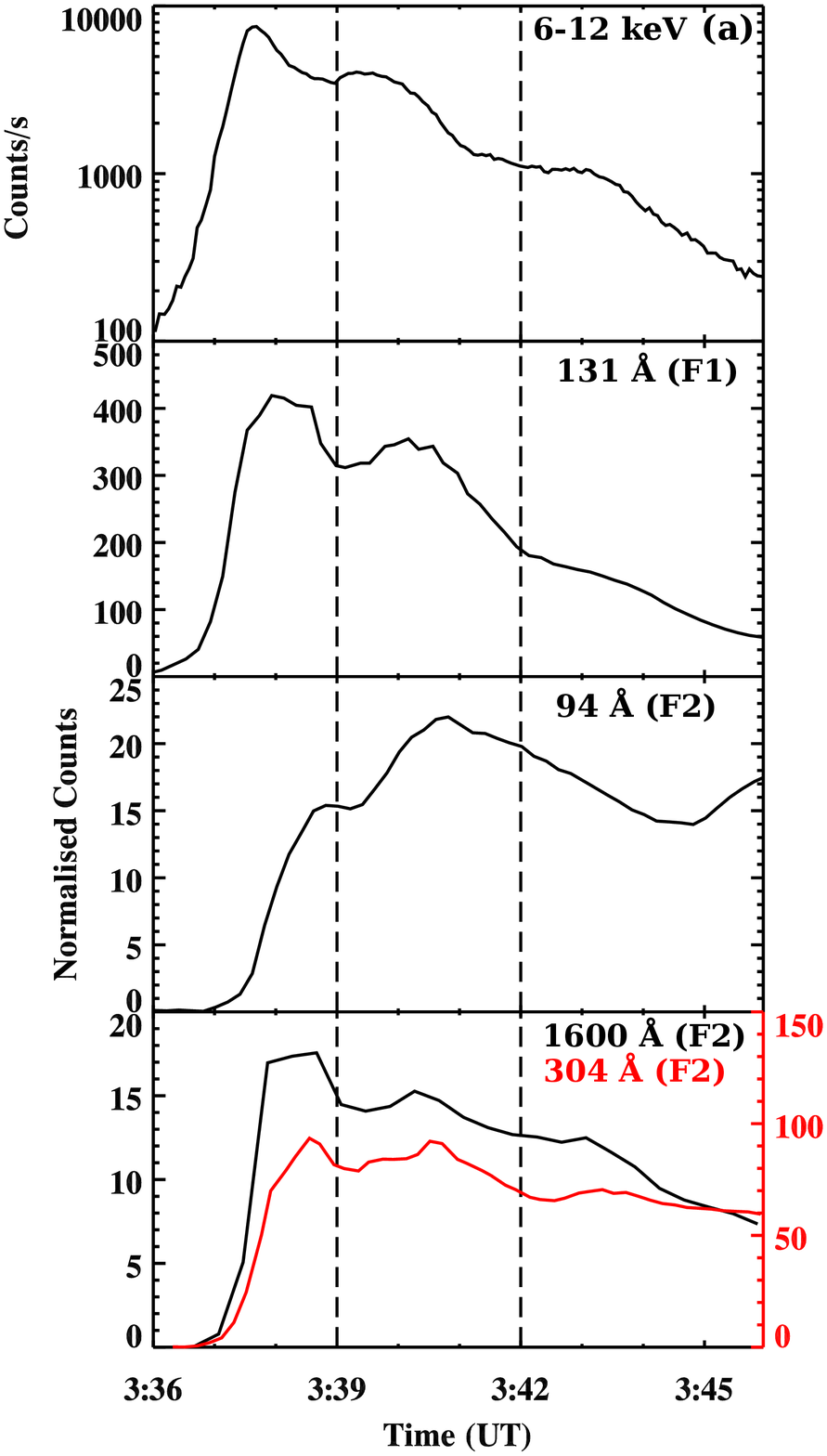}
\vspace{0.005cm}
\includegraphics[width=7.2cm]{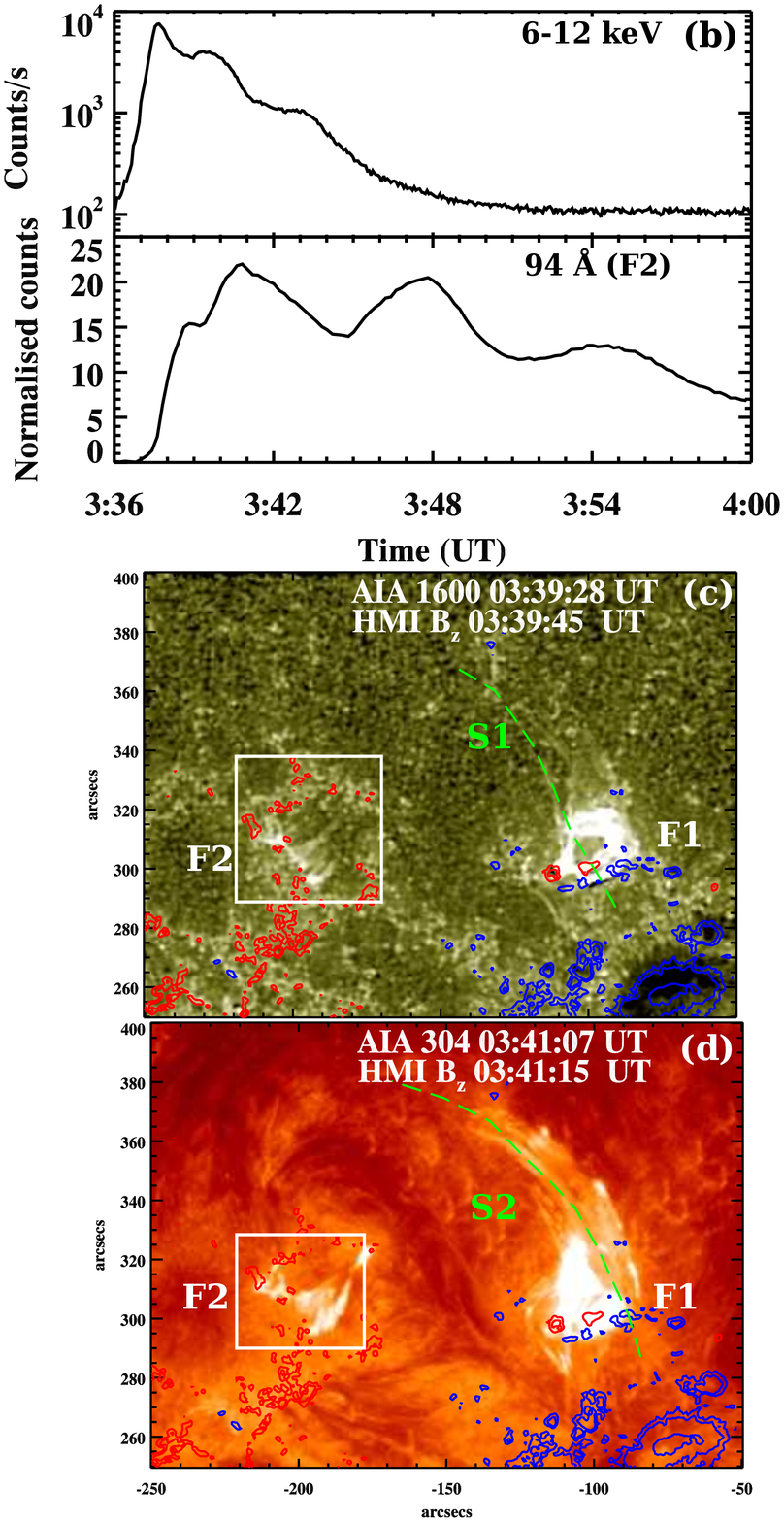}
}
\caption{(a) Fermi GBM X-ray flux profile in the 6--12 keV channel, AIA 131~\AA~ base-difference average intensity profile at the footpoint F1, AIA 94, 1600 and 304~\AA~ base-difference average intensity at the footpoint F2. (b) Time profiles of the decaying oscillation in 6-12 keV and AIA 94 \AA~ channel (at footpoint F2). (c-d) AIA 1600 and 304~\AA~ images showing the box region at footpoint F2 where we extracted the normalized intensity using base-difference images. These images have been overlaid by co-temporal HMI magnetogram contours of the positive (red) and negative (blue) polarities. S1 and S2 represent the slice cuts which are used to create the stack plot along the selected path (green dashed line). }
\label{aia1600_304}
\end{figure*}

Figure \ref{aia1600_304}a displays the flux profiles in 6--12~keV, 131, 94, 1600 and 304~\AA~ channels. The 131~\AA~ channel is sensitive to the hot plasma related to the magnetic reconnection region. Fortunately, 131~\AA~ channel images at the footpoint F1 are not saturated during the flare impulsive phase, therefore, we observe clearly the similar decaying oscillation as in 6--12~keV. AIA movies (aia94$\_$bd.mp4, aia1600.mp4) also show the periodic perturbation of the intensity at F1. On the other hand, AIA 1600 and 304~\AA~ channels are sensitive to the photospheric, chromospheric and transition region's temperatures. It is obvious that the magnetic reconnection took place at the footpoint F1, and we see the remote ribbon brightening at the footpoint F2 in the AIA 1600 and 304~\AA~ channels. In addition, there is a magnetic connection between the footpoints F1 and F2. Therefore, particles are accelerated from the reconnection site (at F1) and transported to the other footpoint (F2) along the arcade loops during the periodic magnetic reconnection. Thus, the normalized fluxes at the footpoint F2 in 1600 and 304~\AA~ channels (Figure \ref{aia1600_304} c, d) are consistent (showing three peaks) with the  6--12 keV emission. Therefore, it is clear that the repeated periodic particle acceleration occurred at the reconnection site (footpoint F1). The electrons precipitated into the chromosphere at the footpoint F2 and caused the remote ribbon brightening.  Figure \ref{aia1600_304}b demonstrates the comparison of the two oscillation profiles, in 6--12 keV and AIA 94~\AA~ channels. The onset of the periodic disturbance in 94~\AA~ is closely related with the impulsive energy release at the footpoint F1.


\begin{figure*}
\centering{
\includegraphics[width=9.0cm]{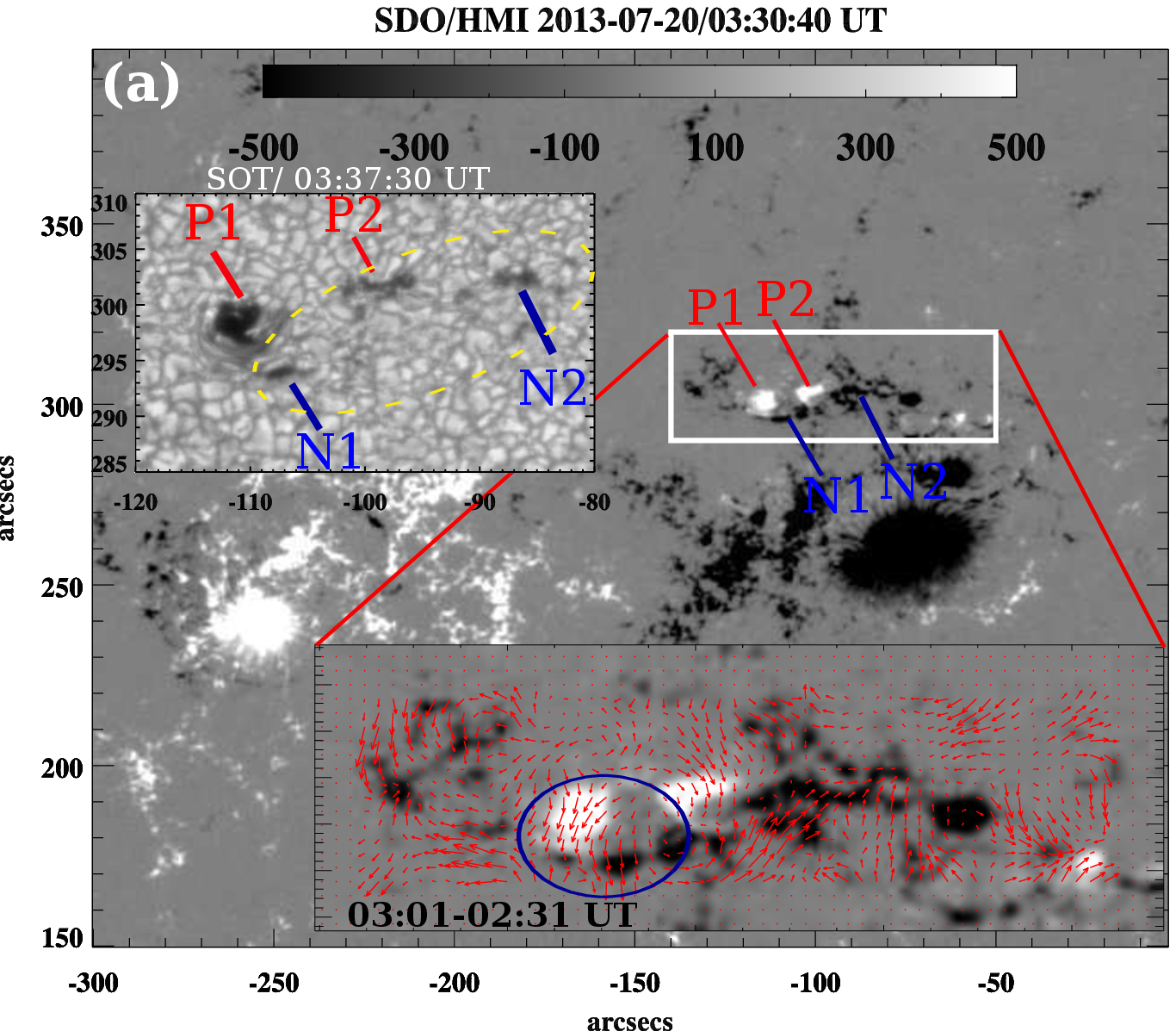}
\includegraphics[width=5.5cm]{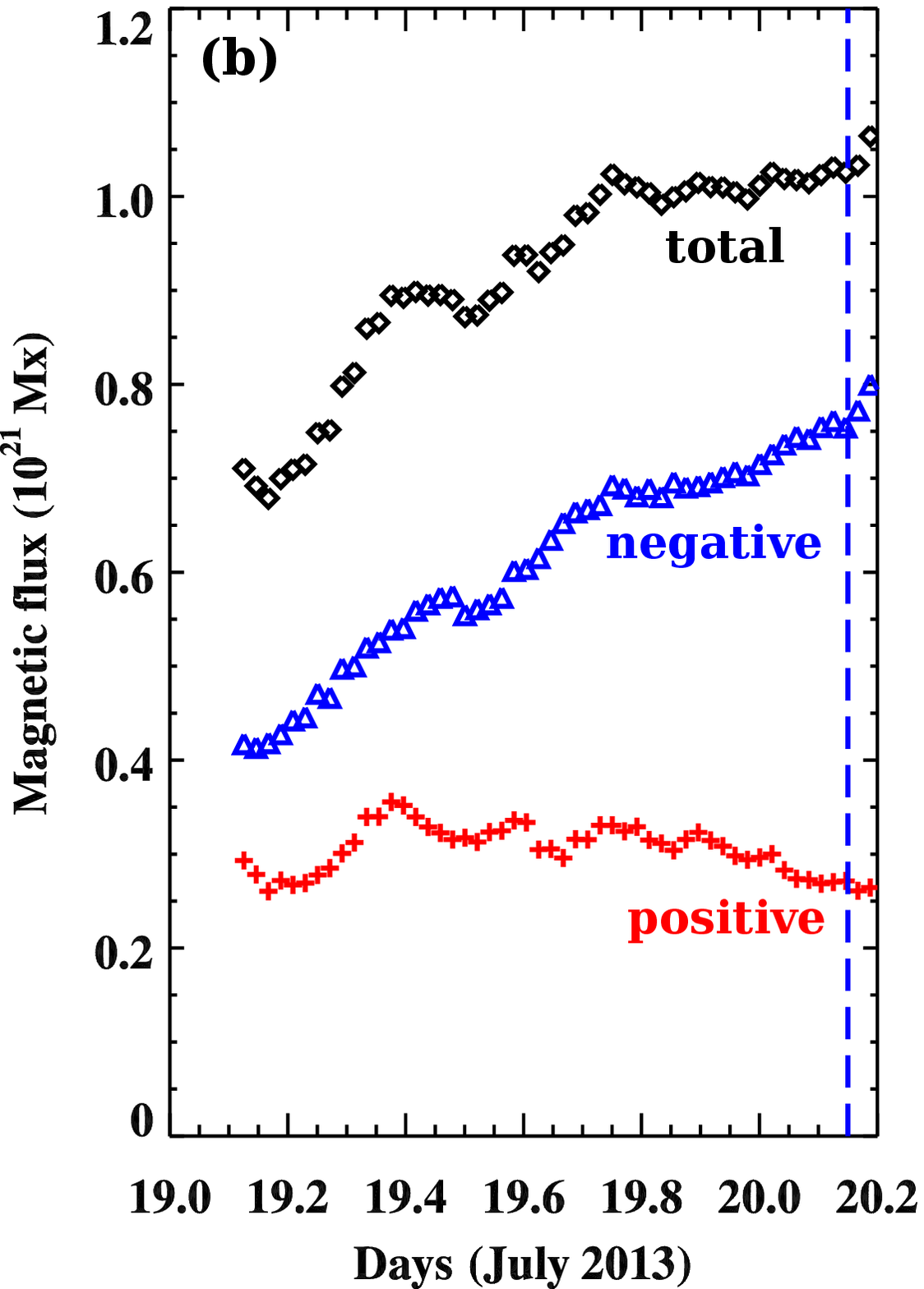}

\includegraphics[width=7.5cm]{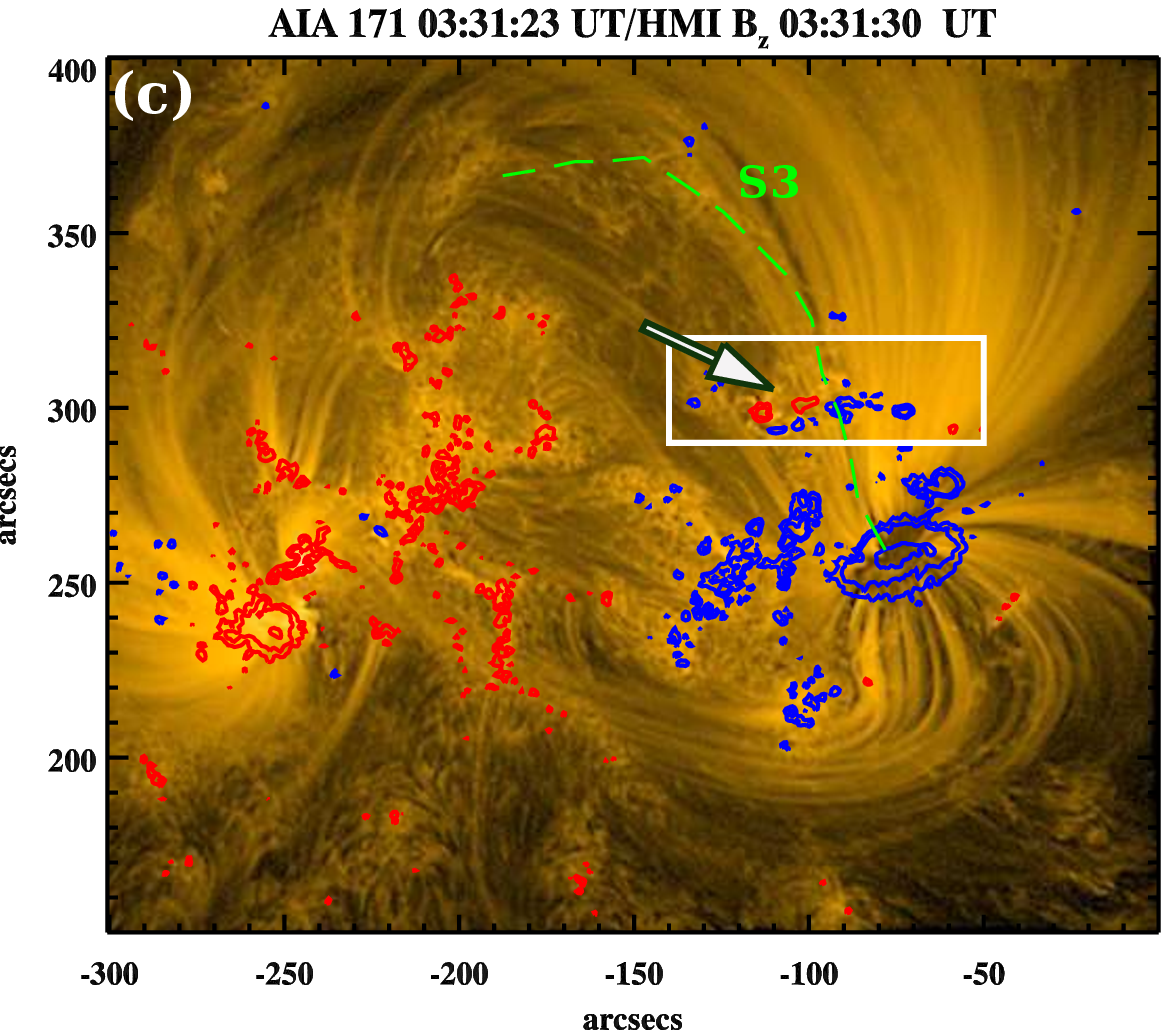}
\includegraphics[width=7.5cm]{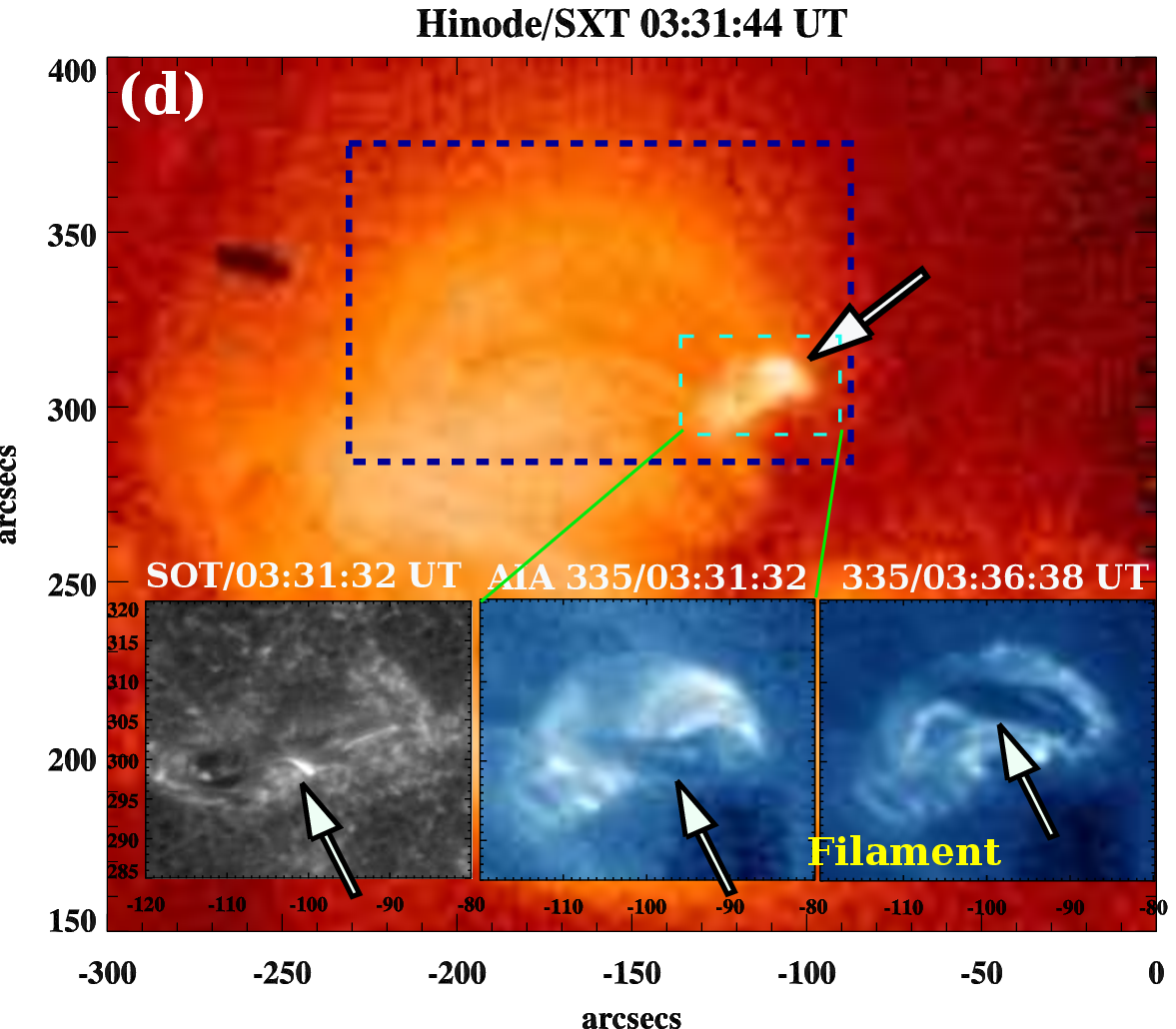}

\includegraphics[width=7.5cm]{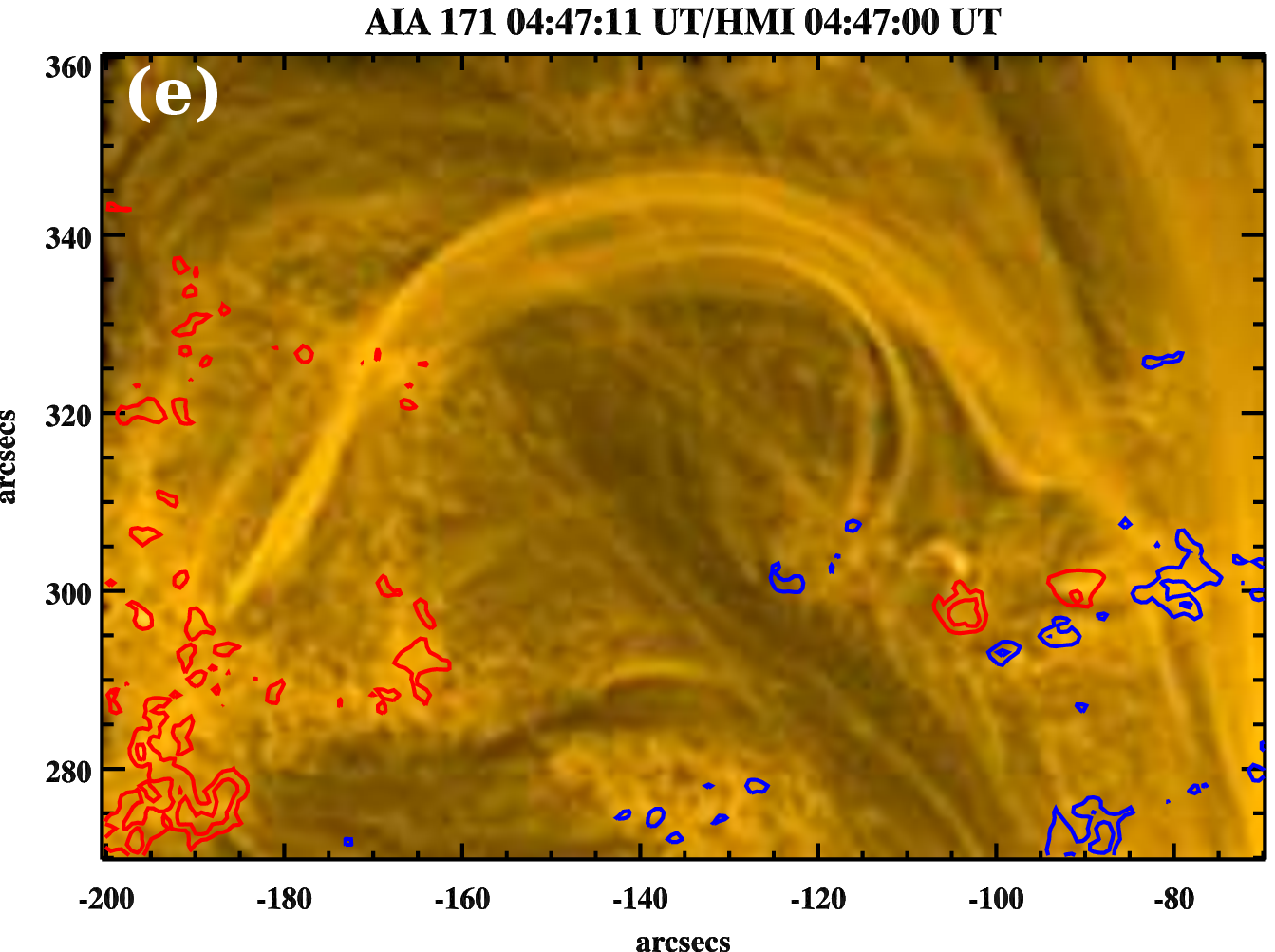}
\includegraphics[width=7.5cm]{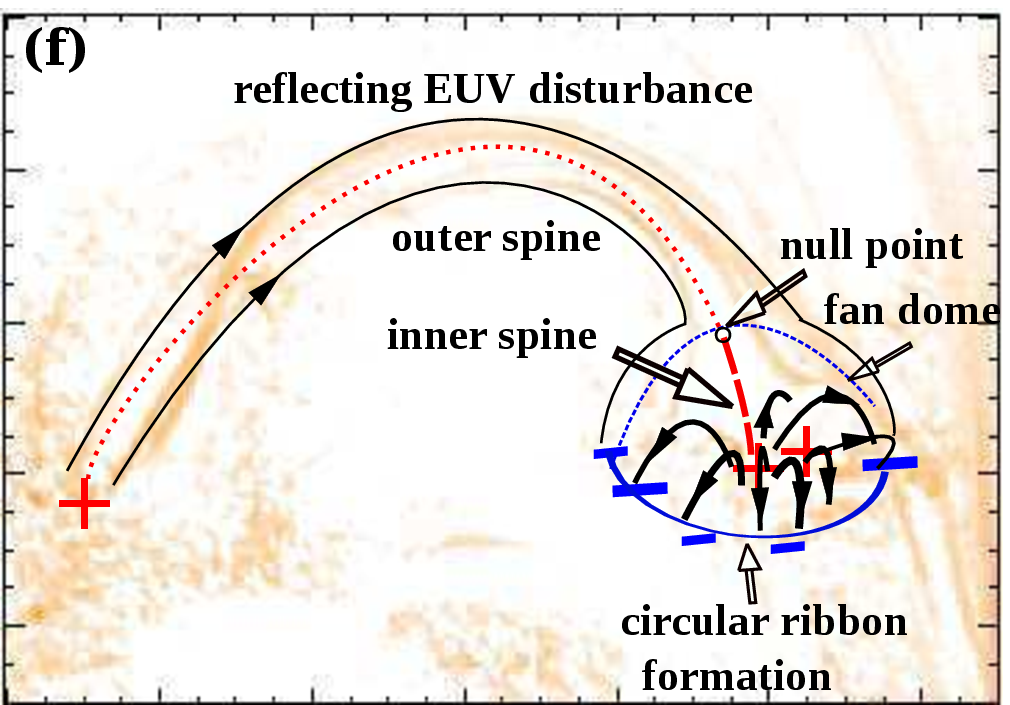}
}
\caption{(a) HMI magnetograms of the active region NOAA 11793 showing the emergence of the new magnetic flux at the flare site (within the box). The photospheric flow map of the selected box area. The longest arrow corresponds to the flow speed of $\sim$442~m~s$^{-1}$. The inset shows Hindode/SOT blue-continuum image (4504~\AA) of the flare region showing the closer view of the sunspots. P1, P2, N1, N2 indicate the positive and negative polarties sunspots, respectively. (b)  Positive (red), unsigned negative (blue) and unsigned total (black) magnetic flux profiles (in Mx) within the selected box area (before and after the flare) during 19--20 July 2013. The vertical dotted line indicates the flare peak time. (c, d) AIA 171~\AA~ and Hinode XRT images of the AR before the flare onset. The emerging fan or \lq\lq anemone\rq\rq\ loops are marked by arrows. The inset shows the enlarged view of the emerging fan loops and the filament (marked by arrow) observed in the SOT/\ion{Ca}{2} H (3968~\AA) and AIA 335~\AA\ images before the flare onset (at 3:31~UT and 03:36~UT). S3 represents the slice cut which is used to create the stack plot along the selected path (green dashed line). (e) AIA 171~\AA~ after the flare showing the fan-spine topology during the cooling of loops. (f) The schematic cartoon of the fan-spine topology.}
\label{hmi}
\end{figure*}


\subsection{Magnetic configuration and trigger of the oscillations}

To investigate the magnetic configuration of the flare site, we used HMI magnetograms. The HMI movie (hmi.mp4) shows the flux emergence of both positive and negative polarities about 10 hrs before the flare. Figure \ref{hmi}a shows the HMI magnetogram of the AR at 03:30:40 UT (before the flare), where the box area represents the flare site. The inset shows the photospheric flow map of the selected region which is derived using the Differential Affine Velocity Estimator (DAVE) method \citep{schuck2006}. We used a 30-minute time difference between the two coaligned magnetograms. The longest arrow represents the flow speed of 442~m~s$^{-1}$. After the flux emergence, we note continuous shear flows of the positive polarities (within the circle) fields (within the box, Figure \ref{hmi}a) in the counterclockwise direction. The continuous shear flows help in the build-up of magnetic energy at the flare site. 

To explore the emerging mini-sunspot structures, we used  high-resolution (0.1$\arcsec$ per pixel) Hinode/SOT (Solar Optical Telescope; \citealt{tsuneta2008}) blue-continuum (4504~\AA, photosphere) and \ion{Ca}{2} H line (3958~\AA, lower chromosphere) images (1k$\times$1k) of the flare site.  Figure~\ref{hmi}a shows the SOT/blue continuum image of the emerging small region. P1, P2, N1, and N2 are the positive and negative polarity sunspots. The main spot P1 emerged about one day before of the flare onset, which showed  continuous counterclockwise rotation. Later P2, N1, and N2 emerged around P1. Initially P1 emerged as a pore (without penumbra) one day before, and later right-handed twisted penumbral filaments are observed before the flare. This indicates the accumulation of twist by rotation of P1 and formation of twisted PIL (polarity inversion line) after the emergence of P2, N1 and N2. 

Figure~\ref{hmi}b shows the positive, the absolute value of negative, and total magnetic flux profiles within the selected box in the HMI magnetogram. We can see the considerable amount of flux emergence of opposite polarities. The vertical line shows the peak time of the flare. The new flux emergence and shear motion both may be responsible for the flare trigger. 
 On the other hand, we do not observe noticeable flux cancellation at the flare site. New flux emergence within the existing active region, and the rotation of the opposite polarity sunspots (forming \lq\lq anemone\rq\rq\ loops) are most likely evidences of the emergence of a twisted flux rope from under the photosphere \citep{fan2009}, and was found to be associated with flares/CMEs (e.g., \citealt{kumar2013s}).

Figures~\ref{hmi}c,d display the AIA 171 \AA~ and Hinode X-ray telescope (XRT) \citep{golub2007} images of the AR before the flare. The AIA 171~\AA~ channel samples cooler loops ($T \approx 0.7$~MK), whereas XRT (the Ti-poly filter) image shows the hot plasma structures (peak temperature response $\sim$10~MK). The emerging fan loops are marked by an arrow. The closer views of the emerging fan loops are shown in the SOT/\ion{Ca}{2} H line and AIA 335~\AA\ images taken before the flare (Figure~\ref{hmi}d). The \ion{Ca}{2} H line image (at 03:31~UT) shows P1 and the twisted polarity inversion (PIL) line. A small filament lies along the PIL. The AIA 335~\AA\ image at 03:31~UT shows the emerging fan loops and a dark filament lying along the PIL. This filament slowly rises and pushes the fan loops upwards just before the flare at 03:36~UT. The magnetic configuration is quite similar to the fan-spine topology. The XRT image clearly shows the fan and outer spine/connecting loops to the opposite polarity regions. Interestingly, we notice small brightening within the fan loops in the XRT image, which may be due to the build-up of currents. Note that the outer spine and fan connecting loops to the opposite polarity regions are not observed in the 171~\AA~ channel (before the flare). During the magnetic reconnection, the fan rotates in the clockwise direction, (i.e., relaxing the shear) which results in the formation of the circular ribbon at F1. The remote ribbon brightening at F2 moves in the clockwise direction. 
After the flare, the fan-spine loops cool down and become visible in the 171~\AA~ channel (post reconnection configuration) at 04:47~UT (Figure~\ref{hmi}e). Figure \ref{hmi}f shows a schematic cartoon of the possible magnetic configuration derived from the observations. The inner spine, outer spine, null-point and fan are marked by arrows.

\begin{figure*}
\centering{
\includegraphics[width=12.0cm]{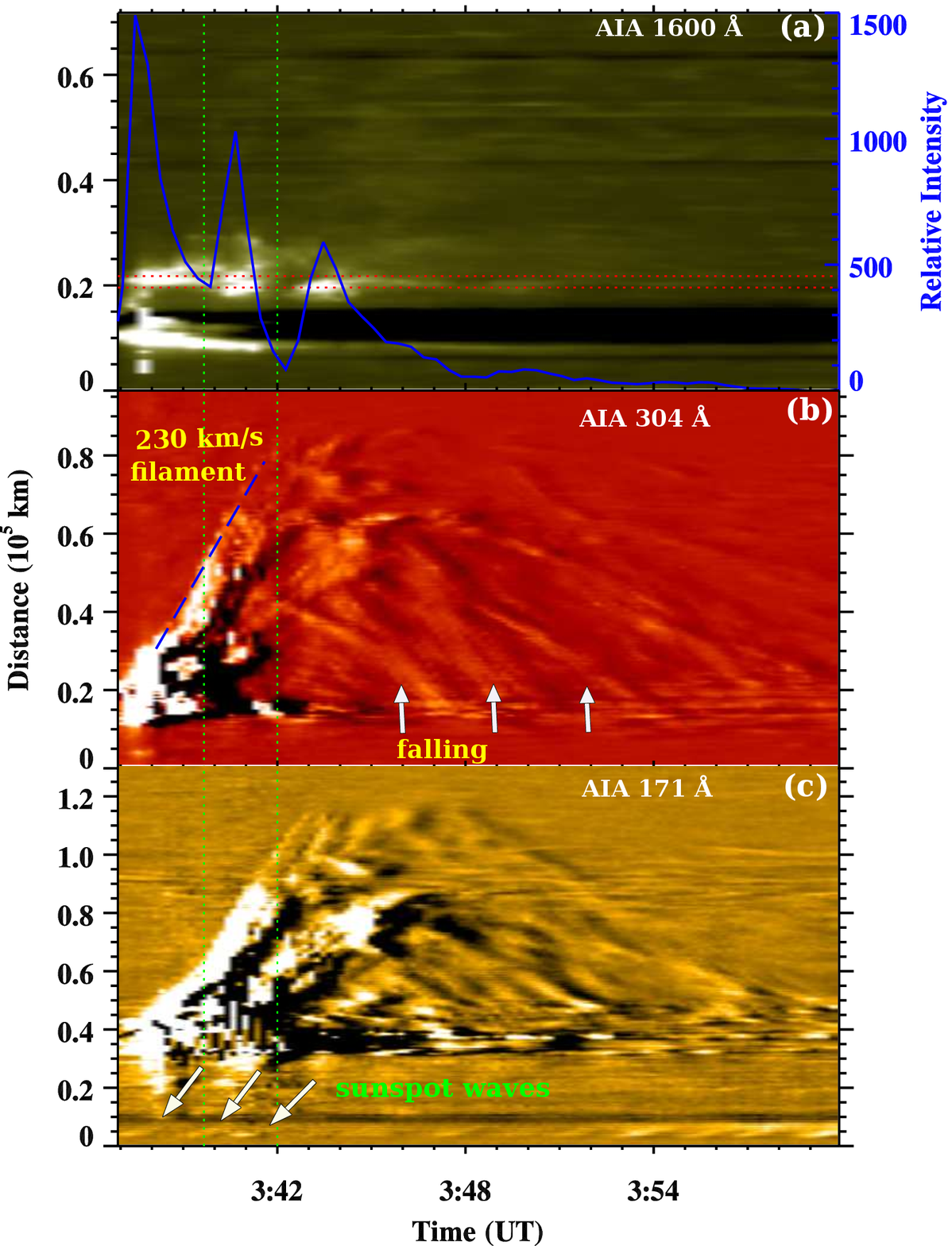}
}
\caption{Time-distance plots along slices S1, S2 and S3 using AIA 1600~\AA~, 304~\AA~ (base difference) and 171~\AA~ (running difference)  images, respectively. The blue curve is the AIA 1600~\AA~ average intensity between two horizontal red dotted lines. The AIA 304 and 171~\AA~ plots show the  filament rising up and then falling back to the solar surface. The mean speed of the filament is 230~km~s$^{-1}$. AIA 171~\AA~ plot shows the sunspot waves (3-min period) emanating from a nearby big sunspot.}
\label{stack_3}
\end{figure*}


To investigate the trigger of the repeated energy release at the null-point, we used AIA 1600, 304, and 171~\AA\ channel images. 
Figure \ref{stack_3}a displays a stack plot of intensity distribution along slices S1, S2, and S3 shown in the AIA 1600~\AA, 304~\AA\  (base-difference) and 171~\AA\  (running difference) images (see Figures~\ref{aia1600_304}c,d and \ref{hmi}c). The AIA 1600~\AA~ average base-difference intensity between two horizontal dotted lines (red) is plotted on the right vertical axis. The decaying oscillation is very clear at a projected (sky plane) height of about 20~Mm (blue curve), which is consistent with the X-ray flux in 6--12~keV. Moreover, we see the continuation of the oscillation after 03:48~UT in 1600~\AA. An AIA 1600~\AA~ movie (aia1600.mp4) also reveals the periodic brightening (apart from the circular ribbon) associated with the rising motion of the filament channel. If we compare it with the XRT and AIA 171~\AA~ images with 1600~\AA~ intensity enhancement, it looks like that it emanates most likely from the null-point, where the periodic reconnection occurred. The first impulsive peak of the flare was associated with the rising of a filament below the emerging fan (AIA 304, 335, and 171~\AA). Other energy releases are closely associated with the untwisting motion of the filament. Later, the filament starts falling back to the solar surface. Probably overlying magnetic field does not allow it to escape, i.e., the phenomenon of failed eruption occurs in this event (e.g., \citealt{kumar2011}). We assume from the stack plot that the uncertainty in the distance measurement is 5 pixels.
The mean speed of the rising filament, estimated from a linear fit, was 230$\pm$15~km~s$^{-1}$. The AIA 304~\AA~ stack plot (along S2) also shows the brightenings associated with the untwisting filament. Moreover, the sunspot waves (of a 3-min period) emanating from a nearby big sunspot are seen in the AIA 171~\AA~ stack plot. 
It is obvious that the periodic energy release is associated with the rising filament. A careful investigation of the movies (1600, 304 and 171~\AA, i.e., aia1600.mp4, aia304$\_$171$\_$int.mp4) suggests that there was a periodic brightening (i.e., intensity perturbation) above the circular ribbon, which was cospatial with the null-point. The energy release most likely took place at the null-point of the fan-spine topology. It seems that the untwisting motion of the filament is associated with the periodic energy release at the apparent location of the null-point.

Low energy QPP (6--12~keV) and longitudinal oscillations seem to be independent. Longitudinal oscillations are triggered by the impulsive energy release (followed by untwisting and rising filament) at F1. At footpoint F2, chromospheric plasma heating is caused by the periodic precipitation of accelerated particles travelling along the hot arcade loops. At F2, the longitudinal wave seen in 94~\AA~ does not show correlation with the AIA 304 and 1600~\AA~ fluxes. Moreover, the returning wave, after bouncing from footpoint F2 does not trigger the repeative reconnection at the null-point (located at F1). 

The short period (202~s) oscillation in EUV and X-ray channels is most likely caused by the periodic particle acceleration in the vicinity of the footpoint F1 towards F2, along the hot loop. The long period oscillation (409~s) in EUV intensity (distance-time plot along the loop) is caused by the compressive slow-mode wave propagating back and forth along the hot loop. A careful comparison of the flux profiles (EUV and X-ray) and images suggests that these oscillations are independent of each other (i.e., the X-ray oscillation is not caused by the reflecting slow-mode wave in the loop). A detailed discussion of this interpretation is given in the next section. In addition, such slow mode oscillations often repeat as a separate set of distinct events as shown by \citet{curdt2003}, and color of the initial Doppler shift (blue or red) is always the same. However, our study demonstrates only a single event, i.e., one distinct energy-release near one footpoint followed by a strongly damped oscillation.

\section{SUMMARY AND DISCUSSION}

 We present a multi-wavelength study of almost simultaneous short and long period decaying oscillations observed during a C-class flare. The period and decay time of the long period oscillation manifested as a wave of the EUV emission intensity periodically bouncing along the arcade loops, are found to be about 409~s and 1,100~s, respectively. The phase speed of the wave was estimated to be about 560~km~s$^{-1}$, which is roughly consistent with the sound speed at temperature 10--15~MK. These long-period oscillations are consistent with the SUMER Doppler-shift oscillations observed in hot coronal loops, which are interpreted as standing slow magnetoacoustic waves reflecting back and forth between the footpoints of the arcade loops \citep{wang2011,ofman2012}. We need to point out that in our case the oscillation is seen as the variation of the emission intensity, in contrast with the classical manifestation of SUMER oscillations in the Doppler shift. A similar oscillation of the plasma density observed in the microwave band was recently reported by \citet{kim2012}. Our observations are also consistent with the standing (bouncing between the footpoints) longitudinal wave, which was directly observed in a different event by \citet{kumar2013w}. The oscillation periods observed by SUMER are typically significantly longer (7--31 minutes) than those presented here ($\sim$6 minute, close to the lower limit), which can be explained by the different observing geometry, i.e., a shorter loop length. Thus, we conclude that the 409-s oscillation belongs to the class of SUMER oscillations and hence can be interpreted as the global standing slow magnetoacoustic wave excited by an impulsive energy release.

The shorter period oscillation, with the period of about 202~s and the decay time of about 154~s, was observed almost simultaneously with the longer period (409~s) oscillation. The 202-s QPP observed in the X-ray channel is mainly originating from footpoint F1. However, same QPP is observed at footpoint F2 in the EUV channel (AIA 1600, 304~\AA). This QPP is likely to be excited by the periodic variation in the flux of accelerated electrons moving from the reconnection site at F1 along the loop and precipitating near F2. Therefore, we cannot rule out the contribution to the X-ray emission from F2: perhaps, the X-ray emission comes from both footpoints. In addition, the 202-s period variations of the EUV emission originating from F1 (AIA 1600, 131~\AA) and F2 (AIA 1600, 304~\AA) show good correlation.

The shorter, 202-s periodicity is almost two times shorter than the longer, 409-s, periodicity, which suggests some possible relationship between them. In particular, the 202-s oscillation could be produced by another eigenmode of the loop, the second standing slow magnetoacoustic harmonics. In this mode the nodes of the velocity perturbations are at the footpoints, and an additional node is at the loop apex \citep[see, e.g., modelling in][]{nakariakov2004,tsiklauri2004}. This mode produces in-phase variations of the emission at the loop footpoints, which is consistent with our findings. But, against the interpretation of the 202~s periodicity in terms of the second slow harmonics is the lack of the observational evidence of this mode in the EUV emission in the loop legs. On the other hand, the 
202-s periodic variation of the emission associated with non-thermal electrons (e.g., in the X-ray band) may be explained by the periodic triggering the magnetic reconnection or its rate by the second slow mode in the magnetic fan configuration situated near F1. But, in this case it is not clear why similar triggering is not caused by the 409-s mode.

The shorter, 202-s periodicity may also be associated with the variation of the plasma density near one or the other footpoint in the global standing longitudinal oscillation of the longer, 409-s period. In the lack of the spatial resolution, this alternate variation of the density near the footpoints (one or the other) would have the period two times shorter than the period of the global mode. 
However, the emission recorded in 6--12~keV is likely to be associated with non-thermal electrons precipitating at the footpoint(s) of the flaring loop, otherwise the temperature of the plasma should exceed 60~MK, that is not directly affected by the variation of the background plasma density. Moreover, the simultaneous appearance of the 202-s oscillation in both footpoints F1 and F2 contradicts to this interpretation too, as in the global standing slow magnetoacoustic mode the density variations at the opposite footpoints are in anti-phase with each other. Thus we disregard this interpretation.

Another option is that the factor of two difference between the 202-s and 409-s periodicities is just a coincidence, and that the 202-s oscillation is connected with the spontaneous repetitive regime of the magnetic reconnection near footpoint F1. The magnetic configuration of the flare site reveals a fan-spine topology with a null-point. The emerging fan loops are clearly observed in the AIA 304~\AA~ images. The presence of the null-point and outer spine connectivity (before the flare onset) is revealed by Hinode/XRT and AIA 171~\AA~ images. Generally,  3D reconnection involves a separatrix dome, thus forming a circular ribbon (e.g., \citealt{fletcher2001,masson2009,pontin2013}). We see continuous emergence of rotating spots of opposite polarity. This rotation helps to build-up the magnetic shear. This could be a signature of the emergence of a flux rope from below the photosphere \citep{fan2009,kumar2013s}. The emerged flux rope can trigger the magnetic reconnection at the null-point, similar to the magnetic breakout mechanism \citep{antiochos1998,sun2012}. The average speed of the untwisting cool plasma injection was seen to be $\sim$230~km~s$^{-1}$.  In addition to the production of the 202-s QPP, the localised energy release at the null-point most likely launches the 409-s wave that is seen to bounce back and forth along the arcade loops. This scenario is consistent with the 
conclusion of  \citet{wang2003b} who have reported that 50\% of the observed SUMER oscillation events were not associated with a flare, suggesting that the oscillations were induced by a flare at one of the footpoints that was in those cases behind the limb.

Magnetic field extrapolations of circular ribbon flare events revealed the fan-spine connectivity to a remote ribbon in other events, e.g. studied by \citet{masson2009}. However, the hot loops/outer spine connecting to the remote ribbon site was not possible to observe by TRACE, if the arcade loops were heated to more than 6~MK. They can be observed only in the hot channels ($\sim$10~MK) provided by the new generation of solar instruments, either in AIA 131, 94~\AA\ channels of AIA or with Hinode/XRT. 

The observations suggest that the periodically accelerated electrons that apparently cause the 202-s oscillation, propagate from the null-point situated near footpoint F1 to the opposite footpoint, F2. If the outer spine was open to the interplanetary medium (for example in the case of X-ray/EUV jets), the particles might have escaped into the interplanetary medium too. These electron beams would generate repeated (quasi-periodic) type III radio bursts. In our case, we did not find any type III radio burst (Wind/WAVES, STEREO-A and B radio dynamic spectra), which indicates that either the accelerated particles are confined to the arcade loops/outer spine and generate the remote flare ribbon (i.e., towards the footpoint F2) or the flux of the non-thermal electrons going upward is too weak to cause the radio bursts.

Thus we are inclined to conclude that the shorter-period oscillation with the period 202~s is not likely to be connected with the longer-period, 409~s, oscillation. The 202~s oscillation could be associated with periodic modulation of the acceleration or kinematics of the non-thermal electrons produced by the reconnection occurred at the null-point.
The periodic reconnection can be produced by several mechanisms. Some possible scenarios are given below:

(i) The periodic particle acceleration can be generated by spontaneous bursty magnetic reconnection due to tearing of the current-sheet \citep{kliem2000,barta2008}. Multiple plasmoids are formed and ejected along the current sheet, that generates drifting pulsating structures in the decimetric radio band \citep{karlicky2010,kumar2013p}. Periodicities of these pulsations are estimated to be of the order of a few seconds, that is different from the periodicities detected in our study. Moreover, it is also not clear whether this mechanism can generate a decaying oscillation. 
Therefore, we rule out this interpretation of the detected QPP.

(ii) Slow-mode waves leaking from a sunspot may be a possible candidate for generating the QPP of about 3--5 minutes \citep{chen2006}. SOT images show a positive polarity sunspot (P1) with right-handed twisted penumbral filaments. It is likely that 3-min oscillations in sunspot P1 can produce slow-mode waves toward the null-point (fan-spine topology) along the emerging flux tube or filament, which may periodically trigger the magnetic reconnection with the period about 3 minutes (the effect was demonstrated by \citealt{sych2009}). However, we see the sunspot waves emanating from the nearby big sunspot situated about 40$\arcsec$ away from the flare site. Similarly, the slow-mode waves may emanate from P1, and can cause the periodic reconnection. Compressive slow-mode waves may push the field lines together at certain
phases of the oscillation, which may induce magnetic reconnection of oppositely-directed field lines \citep{ning2004} or simply modulate its rate.

(iii) The periodic reconnection may be triggered by fast-mode waves too \citep{nakariakov2006}. Nonlinear fast-mode wave may deform magnetic X-point to generate repetitive reconnection \citep{mclaughlin2012}. But, we did not observe fast EUV wave trains or wavefronts in the AIA channels \citep[e.g., such as detected by][]{liu2011}, and neither kink nor sausage oscillations were detected near the reconnection site. Therefore, this mechanism does not seem to operate in our case. However, there is a possibility that the fast waves are produced by linear coupling with slow waves discussed in item (ii). If they have a sausage symmetry and propagate perpendicularly to the line-of-sight, they do not cause any noticeable perturbation of the EUV intensity and hence remain invisible \citep{gruszecki2012}. 

(iv) Torsional Alfv\'en waves embedded in the filament/flux rope may also trigger the periodic reconnection. Torsional Alfv\'en waves have been detected in the interplanetary flux ropes \citep{gosling2010}. Sunspots rotation (mainly P1) is considered as an evidence of torsional Alfv\'en wave associated with the emerging flux rope \citep{fan2009}. We observed the sunspot rotation (mainly P1 in the counterclockwise direction) and filament rise below the fan loops before the flare. After the first energy release, we see the untwisting motion of the filament at the reconnection site.  It is likely that the untwisting filament periodically triggers magnetic reconnection at the null-point. A torsional Alfv\'en wave (embedded in the flux tube) can periodically change the angle between the field lines of the flux rope that carries the wave and the neighbouring field \citep{ning2004}.

In conclusion, we report a rare observation of almost simultaneous short and long period oscillations associated with the fan-spine magnetic topology in the corona and untwisting motion of the filament. The long period, 409-s oscillation we interpret as a SUMER oscillation, excited by an impulsive energy release near one of the footpoints of the oscillating loops. We speculate that most of the SUMER hot loop oscillation events may be associated with the fan-spine topology (within the active region) too, where the impulsive reconnection at one of the footpoints launches a slow-mode compressive wave that reflects back and forth along the hot arcade loops. The detected continuous flux emergence and sunspot rotation contribute in the energy build-up at the flare site. The short, 202-s period oscillation is likely to be caused by a repetitive regime of magnetic reconnection, and hence should be attributed to flaring QPP. Similar QPP events should be investigated in more detail for understanding the exact origin of these oscillations.

\acknowledgments
We would like to thank the referee for his/her constructive and valuable comments that improved the manuscript considerably.
SDO is a mission for NASA Living With a Star (LWS) program. PK thanks to Prof. Davina Innes for several fruitful discussions. We acknowledge the use of the Fermi Solar Flare Observations facility funded by the Fermi GI program. Hinode is a Japanese mission developed and launched by ISAS/JAXA, with NAOJ as
domestic partner and NASA and STFC (UK) as international partners. It is
operated by these agencies in co-operation with ESA and the NSC (Norway).
This work was supported by the \lq\lq Development of Korea Space Weather Center\rq\rq\ of KASI and the KASI basic research funds.
VMN acknowledges the support from the European Research Council under the \textit{SeismoSun} Research Project No. 321141, STFC consolidated grant ST/L000733/1, and the BK21 plus program through the National Research Foundation funded by the Ministry of Education of Korea. 
\bibliographystyle{apj}
\bibliography{reference}
\clearpage

\end{document}